\documentclass[useAMS,usenatbib]{mn2e}
\usepackage{graphics,epsfig,psfig}
\usepackage[normalem]{ulem}
\usepackage{xcolor,verbatim}
\usepackage[]{inputenc,amssymb,amsmath}
\usepackage{color}

\def \be{\begin{equation}}
\def \ee{\end{equation}}
\newcommand       \ba           {\begin{eqnarray}}
\newcommand       \ea           {\end{eqnarray}}
\def \bea{\begin{eqnarray}}
\def \eea{\end{eqnarray}}

\def\bm#1{\mbox{\boldmath $#1$}}
\newcommand{\comments}[1]{}

\newcommand       \apjl         {ApJL}
\newcommand       \apjs         {ApJS}
\newcommand       \aap          {A\&A}

\definecolor{webgreen}{rgb}{0,.5,0}
\definecolor{webbrown}{rgb}{.6,0,0}
\usepackage[pdfpagelabels]{hyperref}
\hypersetup{%
   colorlinks=true,%
   breaklinks=true,%
   plainpages=false, bookmarksnumbered, bookmarksopen=true,
   bookmarksopenlevel=1,%
   urlcolor=webbrown, linkcolor=blue, citecolor=webgreen,
   }

\title[ICM]{Multiphase gas in the circumgalactic medium: relative role of $t_{\rm cool}/t_{\rm ff}$ and density fluctuations}

\author[P. P. Choudhury, P. Sharma, E. Quataert]
{Prakriti Pal Choudhury$^\ddag$$^\P$, Prateek Sharma$^\S$, Eliot Quataert$^\dag$\\
$^\ddag$ Department of Physics, Indian Institute of Science, Bangalore, India 560012 (prakritic@iisc.ac.in)\\
$^\P$Max Planck Institute for Astrophysics, Garching 85748, Germany \\
$^\S$Department of Physics and Joint Astronomy Program, Indian Institute of Science, Bangalore, India 560012 (prateek@iisc.ac.in) \\
$^\dag$Astronomy Department, Theoretical Astrophysics Center, University of California Berkeley, Berkeley, CA 94720, USA \\(eliot@berkeley.edu)}

\begin{document}
\maketitle

\label{firstpage}
\begin{abstract}
We perform a suite of simulations with realistic gravity and thermal balance in shells to quantify the role of the ratio of cooling time to the free-fall time ($t_{\rm cool}/t_{\rm ff}$) and the amplitude of density perturbations ($\delta \rho/\rho$) in the production of multiphase gas in the circumgalactic medium (CGM). Previous idealized simulations, focussing on small amplitude perturbations in the intracluster medium (ICM), found that cold gas can condense out of the hot ICM in global thermal balance when the background $t_{\rm cool}/t_{\rm ff} \lesssim 10$. Recent observations suggest the presence of cold gas even when the background profiles have somewhat large values of ${t_{\rm cool}/t_{\rm ff}}$. This partly motivates a better understanding of additional factors such as large density perturbations that can enhance the propensity for cooling and condensation even when the background ${t_{\rm cool}/t_{\rm ff}}$ is high. Such large density contrasts can be seeded by galaxy wakes or dense cosmological filaments. From our simulations, we introduce a condensation curve in the $(\delta \rho/\rho)$ - min$(t_{\rm cool}/t_{\rm ff})$ space, that defines the threshold for condensation of multiphase gas in the CGM. We show that this condensation curve corresponds to ${(t_{\rm cool}/t_{\rm ff})}_{\rm blob} \lesssim 10$ applied to the overdense blob instead of the background for which $t_{\rm cool}/t_{\rm ff}$ can be higher. We also study the modification in the condensation curve by varying entropy stratification. Steeper (positive) entropy gradients shift the condensation curve to higher amplitudes of perturbations (i.e., make condensation difficult). A constant entropy core, applicable to the CGM in smaller halos, shows condensation over a larger range of radii as compared to the steeper entropy profiles in the ICM. 
\end{abstract}

\section{Introduction}
The origin and fate of cold gas ($\lesssim 10^4$ K) in dark matter halos is crucial to our understanding of galaxy formation because it provides fuel for star formation, gets expelled by winds/jets and gets recycled into stars (\citealt{tumlinson17}). 
Observations of quasar absorption lines and $\rm Ly \alpha$ emission (\citealt{rauch11}, \citealt{matejek12}, \citealt{bowen16}) suggest that cold gas pervades the circumgalactic medium (CGM) around galaxies within $\lesssim 100$ kpc, 
 at different redshifts. Existing theories predict
that massive dark matter halos contain hot, virialized gas in approximate hydrostatic equilibrium (\citealt{birnboimdekel2003}). Only the halos less massive than $3\times 10^{11}~M_{\odot}$ are expected to have narrow cold cosmological filaments directly feeding the central galaxies (\citealt{dekel2009}). Hence the origin of the diffuse multiphase gas along almost all lines of sight in the CGM is not well understood. 

On the other hand, emission line filaments and molecular gas are present in the cores ($\lesssim$ few $10$s of kpc) of galaxy clusters and groups at low redshifts (\citealt{accept09}, \citealt{osullivan17}). In the absence of cooling
flow signatures in galaxy clusters, it is presumed that feedback from the central supermassive black hole predominantly helps in maintaining
thermal balance in the cores (\citealt{fabian94}). While local thermal instability is a viable mechanism that can generate dense filaments in the ICM \mbox{(\citealt{field65}, \citealt{balbus88}, \citealt{nulsen97})}, 
there is no consensus on the origin of cold gas in the circumgalactic medium. In this work, we perform idealized simulations to explore the relative roles of different physical parameters important for the condensation in the ICM/CGM. Additionally, we investigate if the physical principles, applicable to understand the formation of cold gas in the ICM, can be consistently generalized in the context of the CGM. 

The formation of cold gas via local thermal instability and its relation to cluster observables have been the focus of observations as well as 
simulations of galaxy clusters (\citealt{sharma10}, \citealt{mccourt12}, \citealt{salom06}, \citealt{tremblay12}, \citealt{voitnat2015}, \citealt{tremblaynat16}).
According to the models built on this idea, these cold clumps provide fuel for the central black hole and feedback jets that maintain 
thermal balance, thus completing the feedback cycle (\citealt{ps05}, \cite{libryan14},\cite{prasad15},\citealt{voit2015}). Earlier simulations have shown that 
condensation due to local thermal instability is triggered only when the background ICM has the minimum ratio of the cooling time to the free-fall
time ($t_{\rm cool}/t_{\rm ff}$) below $\sim 10$ for realistic cool cluster cores (\citealt{mccourt12},  \citealt{sharma12}).

Recently \cite{meece15} claimed, based on their idealized simulations, that cold gas can condense out even if the ratio of the cooling time to the free-fall time is larger than $10$. Following that, \cite{choudhury16}, using global linear stability analysis and idealized simulations, explored
the possibility of higher threshold values of the ratio by considering idealized potentials, somewhat different from the NFW potential typically considered in clusters. 
These studies also find that the $t_{\rm cool}/t_{\rm ff}$ threshold lies around $10$ (within at most a factor of $2$) for realistic clusters, as long as the density perturbations ($\delta \rho/\rho$) are small ($\lesssim 1$). 

\cite{hogan17} recently observed the profiles of $56$ clusters from {\it{Chandra}} X-ray observatory ($33$ out of which are cool cores with H$\alpha$ emission) and deduced that almost all of these have
min$(t_{\rm cool}/t_{\rm ff}) \gtrsim 10$. A larger sample from \citealt{pulido17} shows some cores with min$(t_{\rm cool}/t_{\rm ff}) < 10$, but discrepancies between observations and simulations remain (see section $4.4$ of \citealt{pulido17} for a detailed discussion; see also section 4.1 of \citealt{prasad18}). 
This raises doubts on thermal instability models and gives
impetus to understand multiphase condensation with large $t_{\rm cool}/t_{\rm ff}$.

It is anticipated that large density perturbations make it easy for condensation to occur (\citealt{ps05}, \citealt{ashmeet14}). In this paper, we set up extensive numerical experiments to explore the role of density perturbations in multiphase condensation. We begin with a hydrostatic ICM, confined by the NFW potential (and additionally BCG potential in some cases) and defined by a radially
varying entropy profile, identical to what is described in \cite{choudhury16}. We introduce large, isobaric density inhomogeneities in some of the background profiles. The main motivation of our work is to understand how the initial background $t_{\rm cool}/t_{\rm ff}$ and amplitudes of density perturbations, $\delta\rho/\rho$,  govern the condensation of multiphase gas in the ICM.
Earlier idealized simulations, which found min$(t_{\rm cool}/t_{\rm ff}) \approx 10$ threshold to be 
robust in the cluster-gravity regime, focussed only on low amplitude perturbations ($\delta \rho/\rho$).

We scan the range of the two key parameters [$(\delta \rho/\rho)$-min$(t_{\rm cool}/t_{\rm ff})$] to delineate the space in which multiphase condensation occurs.
We find that large background $t_{\rm cool}/t_{\rm ff}$ requires a large initial perturbation for cold gas to condense out. Large density perturbations may
explain the cold gas in clusters of significantly high background min$(t_{\rm cool}/t_{\rm ff})$ than the threshold value
of $10$. However, we see that the condensation curve roughly traces out the locus of $t_{\rm cool}/t_{\rm ff} \lesssim 10$ for a small overdensity seeded in the background medium. For large amplitudes of perturbations, the effective $t_{\rm cool}/t_{\rm ff}$ of the overdense blob has to fall below a threshold to obtain multiphase condensation in a background medium having a significantly large $t_{\rm cool}/t_{\rm ff}$.  

Radio and X-ray observations show large-scale, low-density cavities that are the relics of jet events in cool-core clusters. The trails of these high-speed
jets can contain tiny regions, roughly at the same pressure with the surroundings, but with a larger local density than that of the background. Larger density perturbations are also expected for lower mass CGM 
where feedback is expected to cause a larger deviation from HSE (\citealt{oppenheimer18}, \citealt{2017MNRAS_fielding}). Cold mode accretion along cosmological filaments can generate large overdensities as well (\citealt{keres09}). We mimic
such large density enhancements in the ICM by putting large amplitude, isobaric density perturbations in our idealized model. 
We also test the robustness of our condensation curve
for various scenarios such as localized perturbations, rising buoyant bubbles, jets, and different entropy stratification. 

\cite{voitglobal17} argued that the role of the radial entropy gradient in the ICM is understated and it primarily regulates
the stochastic cold accretion and feedback cycles. So we study the effects of entropy variation on the condensation curve by using different initial
entropy profiles. We find that the presence of the internal gravity waves with a positive entropy gradient does make condensation difficult, requiring moderately lower values of min$(t_{\rm cool}/t_{\rm ff})$ for condensation. On the other hand with a constant entropy in which internal gravity waves are absent, it is only slightly easier to condense. 
Thus, just the linear response may not predict cold gas condensation very accurately. However, we find disrupted large cores and relatively enhanced cold gas in a constant entropy medium, once condensation is triggered. This flatter entropy profile (large core) is probably more relevant for lower mass halos (such as Milky Way) in contrast to clusters.


In section \ref{sec:simset}, we describe the set up and relevant details of initialization of the virialized ICM. Section \ref{sec:result}
shows the results. We discuss the astrophysical implications of our results in section \ref{sec:disc}. We conclude in section \ref{sec:last}.
\begin{table*}
 \caption{Numerical experiments to quantify the relative role of min$(t_{\rm TI}/t_{\rm ff})$ and $\delta$}
{\centering
\begin{tabular}{c c c c}
\hline\hline
Gravity & Density perturbations & Parameters varied & Additional factors \\

\hline
 NFW & throughout medium & $K_0$, $\delta$ & - \\
 NFW+BCG & throughout medium & $K_0$, $\delta$ & -   \\
 NFW+BCG & localized shell & $K_0$, $\delta$ & -  \\
 NFW+BCG & throughout medium & $K_0$, $\delta$, $K_{100}=0$ & -  \\
 NFW+BCG & throughout medium & $K_{100}$, $\delta$, $\alpha$, $K_{0}=0$ & -  \\
 NFW+BCG & throughout medium & $K_0$, $\delta$, $\dot{M}_{\rm acc}$ & AGN jet \\
 NFW+BCG & throughout medium & $K_0$, $\delta$ & bubble  \\
\hline
\end{tabular}
} \\
\textbf{Notes:} Each of the cases has multiple runs with different values of the parameters.
 \label{table:listofruns}
\end{table*}

\section{Physical/Simulation setup}

\label{sec:simset}
\subsection{Model \& equations}
\label{sec:sec1}
We model the initial background ICM to be in hydrostatic and thermal equilibrium. The following equations describe the evolution of the gas.
\ba
\label{eq:eq1}
\frac{D \rho}{Dt} &=& - \rho {\bf \nabla \cdot \bm{v}} + S_{\rho}, \\
\label{eq:eq2}
\frac{D \bm{v}}{Dt} &=& -\frac{1}{\rho} {\bf \nabla} p - g \bm{\hat{r}} + \frac{1}{\rho}S_{\rho}v_{\rm jet}\hat{r}, \\
\label{eq:eq3}
\frac{p}{(\gamma-1)} \frac{D}{Dt} \left [ \ln \left ( \frac{p}{\rho^\gamma} \right ) \right ] &=& -q^-(n,T) + q^+(r,t), 
\ea

where $D/Dt$ is the Lagrangian derivative, $\rho$, $\bm{v}$ and $p$ are mass density, velocity and pressure; $\gamma=5/3$ is the adiabatic index; 
$q^-(n,T) \equiv n_e n_i \Lambda(T)$ 
($n_e \equiv \rho/[\mu_e m_p]$ and $n_i \equiv \rho/[\mu_i m_p] $ are electron and ion number densities, respectively; $\mu_e=1.17$, $\mu_i=1.32$, 
and $m_p$ is proton mass) 
and $q^+(r,t) \equiv \langle q^- \rangle$ (which imposes thermal balance in shells), $\Lambda(T)$ is the temperature-dependent cooling function. 
We use a fit to the plasma cooling function with a third of the solar metallicity, given by Eq. 12 and the solid line in Fig. 1 of \citet{sharma10}.
Thus the setup is very similar to \cite{sharma12}, the main difference being large density perturbations. In some runs, we inject kinetic jets of constant power with jet mass and momentum source terms which we describe in section \ref{sec:jnfwbcg}. 

\subsubsection{Gravitational potential}
\label{sec:gravpot}
For some of our runs we use the standard NFW (\citealt{nfw96}) gravitational potential as given by Eqs. $4$ and $5$ of \cite{choudhury16}.
For some runs, we add a BCG potential to the NFW potential, of the following form,
\ba
\label{eq:phi_BCG}
\Phi_{\rm{BCG}} = {{V_c}}^2 \ln({r/r_0}),
\ea
where $V_c = 350~{\rm km~s}^{-1}$ and $r_0 = 1~{\rm kpc}$. This accounts for the gravity due to the central galaxy which typically dominates within 
$\sim 10$ kpc.

\subsubsection{Equilibrium profile}
\label{sec:eq}
We have a background hydrostatic 
equilibrium which implies $dp_0/dr = -\rho_0 g$, where a subscript `0' refers to equilibrium quantities and the acceleration due to gravity 
$g \equiv d\Phi/dr$ ($\Phi$ is the fixed gravitational potential).
The entropy profile of the ICM in initial hydrostatic equilibrium is specified as (\citealt{accept09})
\be
\label{eq:ent}
K(r) = \frac{T_{\rm keV}}{n_e^{\gamma - 1}} = K_0 + K_{100}\left(\frac{r}{r_{100}}\right)^{\alpha},
\ee
where $r_{100} = 100 \ \rm  kpc$. We vary $K_0$ to obtain background profiles with different 
$t_{\rm cool}/t_{\rm ff}$. There are runs in which we study the effect of entropy variation. For these cases we either have
$K_0 = 0$ or $K_{100}=0$, and hence vary the non-zero parameter.
Additionally, we have runs in which we vary $\alpha$ to obtain a stronger entropy stratification.

\subsubsection{Important timescales}
The thermal instability (TI) time-scale (the inverse of the local exponential growth rate for a constant 
heating rate per unit volume) is relevant for the isobaric modes and is given by,
\be
\label{eq:tTI}
t_{\rm TI} = \frac{\gamma t_{\rm cool}}{(2 - d \ln \Lambda/d \ln T)},
\ee
where
\be
\label{eq:tcool}
t_{\rm cool} = \frac{nk_BT}{(\gamma - 1)n_e n_i \Lambda}.
\ee
We use min$(t_{\rm cool}/t_{\rm ff})$ and min$(t_{\rm TI}/t_{\rm ff})$ interchangeably as the values are almost equal for clusters.
For only free-free cooling (with $\Lambda \propto T^{\frac{1}{2}}$) relevant to clusters, $t_{\rm TI} = (10/9)t_{\rm cool}$ but it may differ in other cases. The free-fall time
\be 
\label{eq:tuff}
t_{\rm ff} = \left( \frac{2r}{g} \right )^{\frac{1}{2}},
\ee
where $g(r)$ is the gravitational acceleration at the radius of interest.

\subsection{Simulation setup}

We use the {\tt ZEUS-MP} code (\citealt{hayes06}) to solve Euler equations with source terms such as heating, cooling, and gravity (Eqs. \ref{eq:eq1}-\ref{eq:eq3}). The initial condition
consists of a hydrostatic equilibrium profile (described in section \ref{sec:sec1}), superposed with isobaric density perturbations.
\subsubsection{Grids and geometry}
All the simulations are done in spherical ($r$, $\theta$, $\phi$) coordinates with a resolution $256\times256~(N_r \times N_\theta)$.
The radial grid in spherical runs is logarithmic, with an equal number of grid points from 1 to 10 kpc and 10 to 100 kpc. We had to perform hundreds of simulations to map out the condensation curve in the $(\delta - t_{\rm TI}/t_{\rm ff})$ space and hence 3D simulations are prohibitively expensive.

\subsubsection{Initial and boundary conditions} 
\label{sec:ini_bound}
We seed isobaric density perturbations (equivalently, entropy perturbations) described in section 4.1.2 of \cite{choudhury16}, except that we vary the amplitudes in the current
work. We label our runs with the maximum value of overdensity $\delta$ where
\ba
\delta(r,\theta,\phi) = \frac{\rho(r,\theta,\phi) - \bar{\rho}(r)}{\bar{\rho}(r)}
\ea
and $\bar{\rho}$ denotes shell-averaged quantity. The overdensity field is identical for all the runs and we simply scale the amplitude for different simulations. The overdensity field 
is expressed as a sum over Fourier modes in a Cartesian basis and therefore its shell-averaged value is non-zero.
We use $\delta_{\rm max}$ and $\delta_{\rm rms}(r)$ in this work extensively, to denote the maximum value of $\delta$ over the entire simulation box and the root-mean-square
$\delta$ in each shell, respectively.

At the outer boundary the electron number density is fixed to be $n_{e,{\rm out}} = 0.00875~{\rm cm}^{-3}$ for most of the runs. For the runs with a constant initial entropy (and the runs presented in section \ref{sec:compentr}) the value is $n_{e,{\rm out}} = 0.035~{\rm cm}^{-3}$. In this case, we start with higher densities in general because simulations with higher outer densities are better behaved numerically.{\footnote {We observe spurious features emerging at the outer boundary for the constant entropy runs with a lower outer density. We do not quite understand the origin of these features, but these are absent for a higher outer density. By visual inspection of density snapshots, we ensure that 
such artificial features are absent in all our runs.}}We perform appropriate comparisons of these runs having high outer densities with the standard runs to verify the robustness of our results. All these runs with higher $n_{e,{\rm out}}$ show threshold $t_{\rm TI}/t_{\rm ff}$ for condensation, which are consistent with the condensation curve.

The boundary conditions 
in the radial direction allow outflow at the inner boundary ($r_{\rm in}=1$ kpc) and inflow at the outer boundary ($r_{\rm out}=100$ kpc), with the density 
and internal energy density fixed to their initial equilibrium values at the outer boundary. The boundary conditions for $\theta$ ($0<\theta\leqslant\pi$) and $\phi$ ($0<\phi\leqslant2\pi$) directions are, 
respectively, reflective/axisymmetric and periodic. 

\subsubsection{Initial negative density}
\label{sec:negden}
We investigate the effects of very large ($\rm{\delta > 1.0}$) amplitude density perturbations in the ICM and while doing so, we find that some grid points develop negative density at ${t=0}$. To solve this problem and to keep the mean background density profile unaltered, we do the following. We add a floor density at the points with negative density. 
We keep track of the gas mass added in a shell and then subtract the shell-averaged density added from the densities at all points in the shell. 
We do this for each radial shell and thus the background density profile remains unchanged. Mathematically,
\ba
\rho_{\rm{sub}} = \frac{\int_{A_{\rm neg}} (\rho_{\rm{floor}} - \rho_{\rm{nd}})r^2 \sin\theta d\theta d\phi}{\int_A r^2 \sin\theta d\theta d\phi}
\ea
where $\rm{\rho_{nd}}$ denotes the density at the grid points where it becomes negative, $A_{\rm neg}$ implies that the integral in the numerator 
is carried out over the grid points with negative densities in a given shell, $\rm{\rho_{sub}}$ is the density that should be calculated and subtracted for each grid point in the shell
and the integral in the denominator is carried out over all the grid points in the shell.
\subsection{Numerical experiments}
\begin{figure*}
\includegraphics[width=0.8\textwidth]{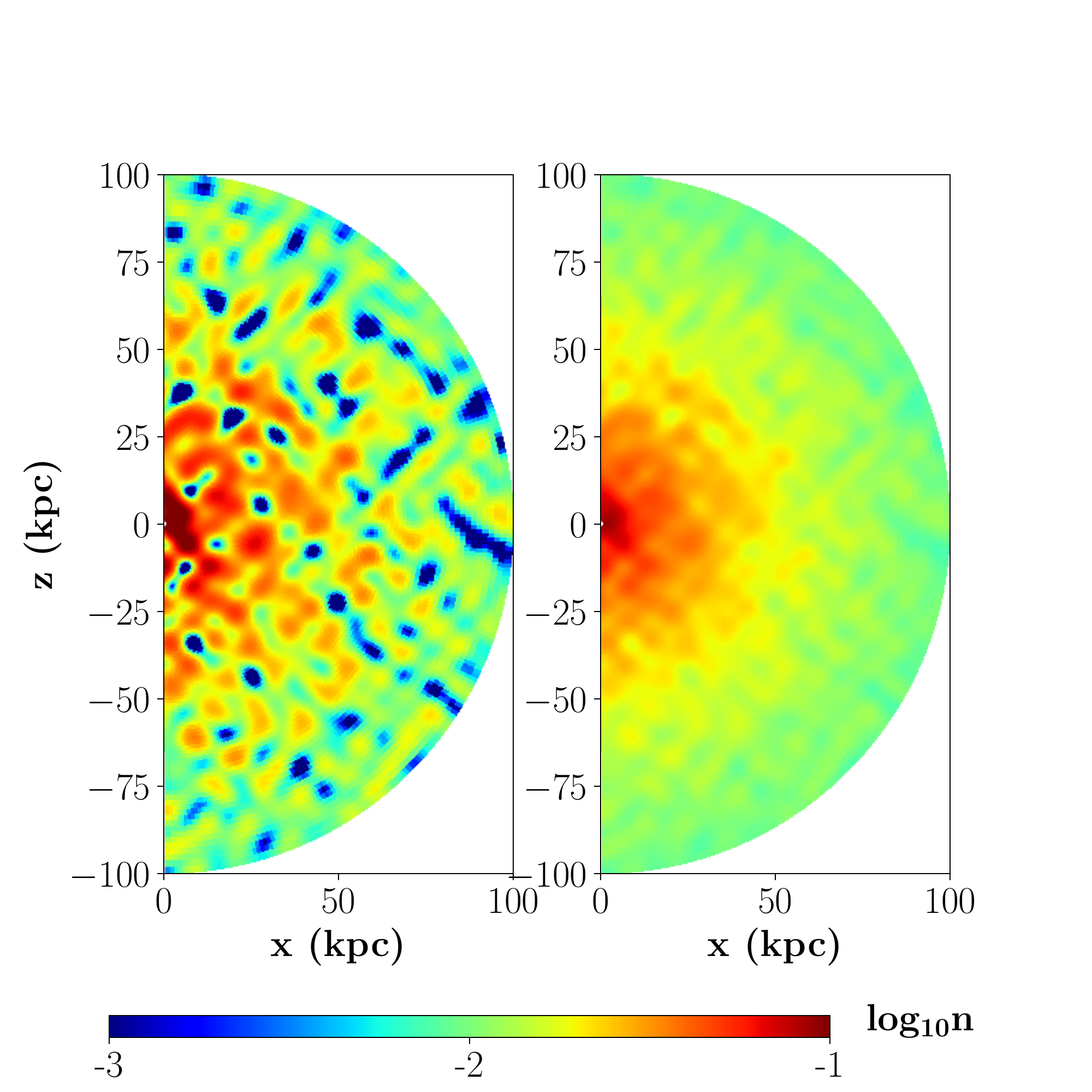}
\caption{The initial number density snapshots for fiducial runs with only NFW gravity $(K_0 = 8~{\rm keVcm}^{2})$ and with large (left) and small (right) amplitudes of density perturbations ($\delta_{\rm max} \approx 1.6$ and $0.32$ respectively). The colorbar is clipped at the maximum and minimum values such that the density larger (smaller) than $0.1$ ($10^{-3}$) $cm^{-3}$ corresponds to the reddest (bluest) color.}\label{fig:fig1}
\end{figure*}

\begin{figure}
\includegraphics[width=0.49\textwidth]{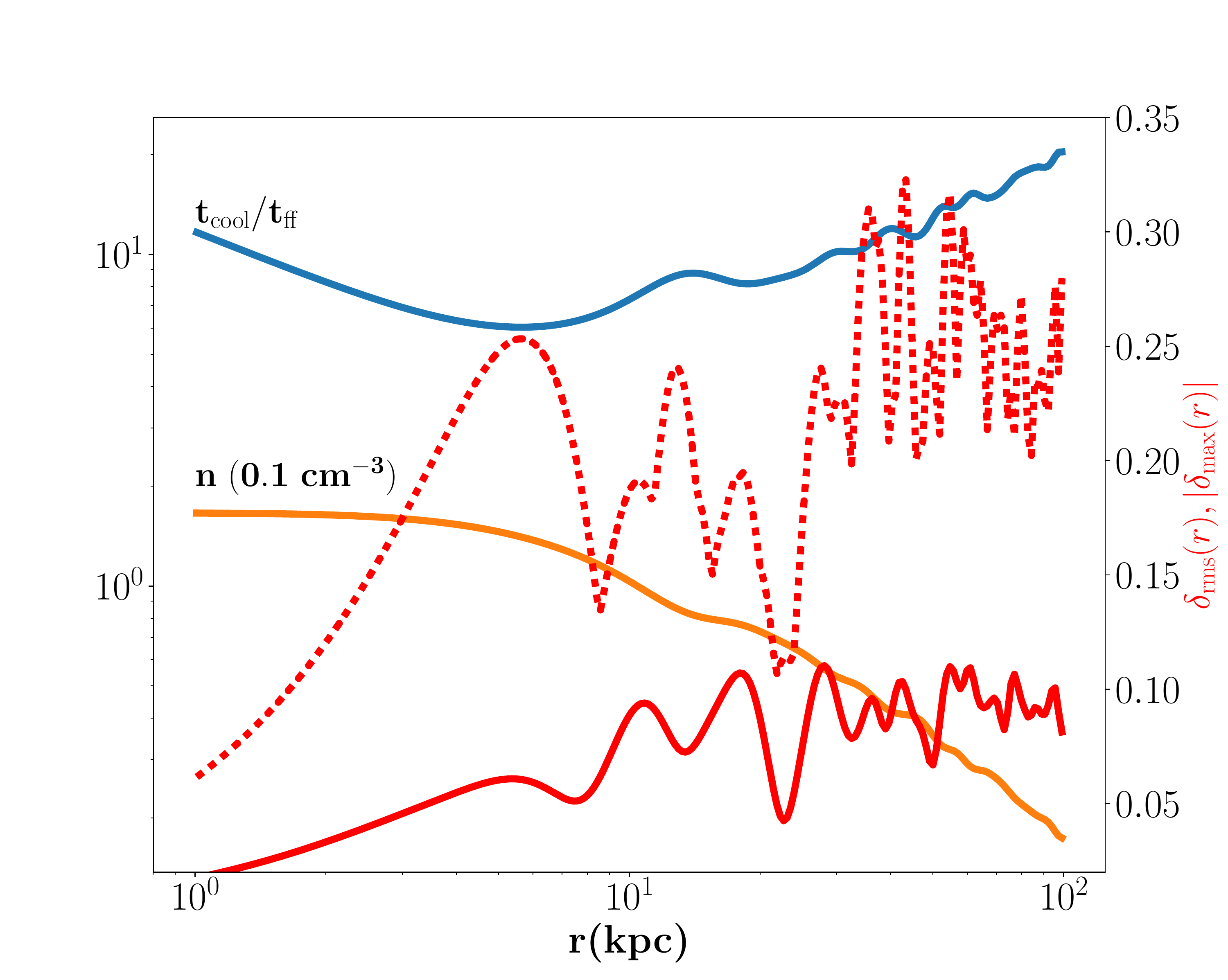}
\caption{The initial shell-averaged profiles with only NFW gravity $(K_0 = 8~{\rm keVcm}^{2})$ and with the amplitude of perturbations $\delta_{\rm max} \approx 0.32$.  The secondary
y-axis represents the scale for density perturbation amplitudes. The solid red line shows $\delta_{\rm rms}$, the rms perturbation (when averaged across $\theta$-direction; see Eq. \ref{eq:eqdelta}) at each radius. The dotted red line shows $\delta_{\rm max}$, the maximum perturbation, at each radius. }\label{fig:fig2}
\end{figure}

In this section we briefly describe the various numerical experiments that we carry out to quantify the relative role of background min$(t_{\rm cool}/t_{\rm ff})$ and
density perturbations in multiphase condensation. Table \ref{table:listofruns} shows all the numerical experiments concisely. 
\subsubsection{NFW potential with different perturbations}
\label{sec:nfw_lowhigh}
For this setup, our gravitational potential is only due to the dark matter halo. The initial background density and cooling time are fixed by the entropy parameters $K_0$ (see section \ref{sec:eq}). Figure \ref{fig:fig1} shows the typical initial conditions for this setup. For different background profiles (with different
min$(t_{\rm cool}/t_{\rm ff})$), we initialize the ICM with different amplitudes of perturbations to map out the regime in which multiphase condensation happens. 
The runs with low amplitude perturbations ($\delta \lesssim 1$) result in the standard min$(t_{\rm cool}/t_{\rm ff})\lesssim 10$ criterion
that was explored most recently in \cite{choudhury16}. For large perturbations ($\delta \gtrsim 1$), we ensure that the lowest density regions
do not have negative values (see section \ref{sec:negden}).

\subsubsection{NFW+BCG potential with different perturbations}
\label{sec:nfwbcg_lowhigh}
For these runs the potential is due to the dark matter as well as the central galaxy (section \ref{sec:gravpot}). 
Recent works (\citealt{hogan17}, \citealt{voitnat2015}, \citealt{prasad18}) have highlighted the importance of including 
the BCG gravity in cool-core clusters.
Here the initialization is done exactly like in section \ref{sec:nfw_lowhigh}. With the free-fall time shorter due
to additional gravity, to get a similar min$(t_{\rm cool}/t_{\rm ff})$, for these runs
we require the cooling times to be lower (or $K_0$ smaller) compared to the NFW-only regime. 
\subsubsection{NFW+BCG potential with localized perturbations}
\label{sec:l_nfwbcg}
In these simulations, we initialize isobaric perturbations only in a few localized spherical shells, with both high and
low amplitudes. We put perturbations within $0.8 H_1$ and $1.2 H_1$ where $H_1$ is the radius around which we intend to perturb the medium.
We want to explore how the radial location of perturbations can shift 
the zone of condensation in the min$(t_{\rm cool}/t_{\rm ff})$ - $\delta\rho/\rho$ parameter space. The first two panels of Figure \ref{fig:fig5} show examples of localized perturbations.
\subsubsection{NFW+BCG potential with perturbations $\&$ a bubble}
\label{sec:bnfwbcg}
In these simulations, we inject a low-density bubble in 2D, of radius $3$ kpc and centered at $4$ kpc so that it touches the inner boundary, and also perturb the background medium similar to some of the previous setups. The bubble has a density of $0.1$ times the background.
The perturbations are small-scale when compared to the spatial extent of the bubble. The bubble will
rise buoyantly and it is interesting to understand how that affects multiphase condensation. The third panel
in Figure \ref{fig:fig5} shows the initialization with a spherical bubble (which is a torus in 3D because of axisymmetry). We also test the results with bubbles centered at $3$ kpc and $2$ kpc with radius $2$ kpc and $1$ kpc respectively. The results are quite similar. 
\subsubsection{NFW+BCG potential with perturbations and entropy variation}
\label{sec:entrv}
We explore how the entropy gradient affects the susceptibility to condensation. There are two sets of runs for this investigation. 
The first in which the initial entropy (see section \ref{sec:eq}) is constant throughout 
the cluster ($K_{100}=0$) and we vary $K_0$. In the other set, we have an initial condition with power-law entropy profiles
($K_{0}=0$) in which we vary $K_{100}$ and $\alpha$. 
\subsubsection{NFW+BCG potential with perturbations and jet}
\label{sec:jnfwbcg}
We inject jets, with mass and radial momentum source terms as in Eqs. \ref{eq:eq1} and \ref{eq:eq2}, of constant mechanical power and mass-loading into the ICM. The source term consists of $S_{\rho} \propto \dot{M}_{\rm jet}$ as described by \cite{prasad15}, where $\dot{M}_{\rm jet}$ is the jet mass-loading factor given by their Eq. $6$ (we use a constant $\dot{M}_{\rm acc}$, $0.01~M_{\odot}/yr$ and $0.1~M_{\odot}/yr$ for two sets of runs, unlike \citealt{prasad15} who calculate $\dot{M}_{\rm acc}$ at $\sim 1~{\rm kpc}$ at each time-step). 
The velocity of the jet is fixed at $\approx 0.2c$ and the jet has an opening angle of $15$ degrees. The spatial distribution is described by
eq. $5$ of \citealt{prasad15}. Strong jets are usually expected to impinge the medium and create low-density regions surrounded by shells of large densities. Hence injection of jets can
cause higher densities, driving condensation for a short time although they heat up the medium on an average and globally reduce the susceptibility to condensation. In our set-up, we also have cooling and heating balanced in shells. Hence injection of jets with large powers causes overheating in the medium and prevents condensation. The two cases of jet injection that we try, inject small mechanical powers, $1.5\times10^{41} {\rm ergs}^{-1}$ (low) and $1.5\times10^{42} {\rm ergs}^{-1}$ (high). Moreover, the low energy injection rates also reflect in a gentle subsonic wind in the cluster core instead of a powerful outflow and the jet does not significantly heat the gas. We prefer to keep the thermal balance in the shells to compare the results with the rest of the cases and understand the effect of jets in multiphase condensation in a background hydrostatic and thermal balance. Hence this setup is not well motivated for jet simulations. Therefore our jet results should only be considered indicative.  

\section{Results}
\label{sec:result}
We explain our results in this section. All our runs are carried out for a maximum of $15$ Gyrs.

The most significant aim of this paper is to delineate 
a condensation zone in the $(\delta \rho/\rho)$ - min$(t_{\rm cool}/t_{\rm ff})$ space. Note that the min$(t_{\rm cool}/t_{\rm ff})$ is measured in the unperturbed ICM. The threshold or the boundary of the condensation zone is simply defined by the min$(t_{\rm cool}/t_{\rm ff})$ or $\delta \rho/\rho$ at which at least some cold gas forms within $15$ Gyrs. We do not quantify the amount of cold gas formed in detail but we have some comparisons of that for the cases with different entropy profiles (section \ref{sec:compentr}). More cold gas is expected the further we go from the threshold line into the condensation zone (\citealt{choudhury16} discuss in more detail
how much gas can condense out for different ICM profiles).  

To assess the regime of multiphase condensation, we vary the density perturbation amplitude characterized by the maximum value of $\delta$, $\delta_{\rm max}$. 
Note that $\delta_{\rm max}$ is the overdensity at a radius, not equal to the radius of min$(t_{\rm cool}/t_{\rm ff})$. It provides an appropriate
label for the amplitude of perturbations of a medium in which inhomogeneities are seeded throughout the medium. 

\subsection{NFW}
\label{sec:resnfw}
\begin{figure}
\includegraphics[width=0.5\textwidth]{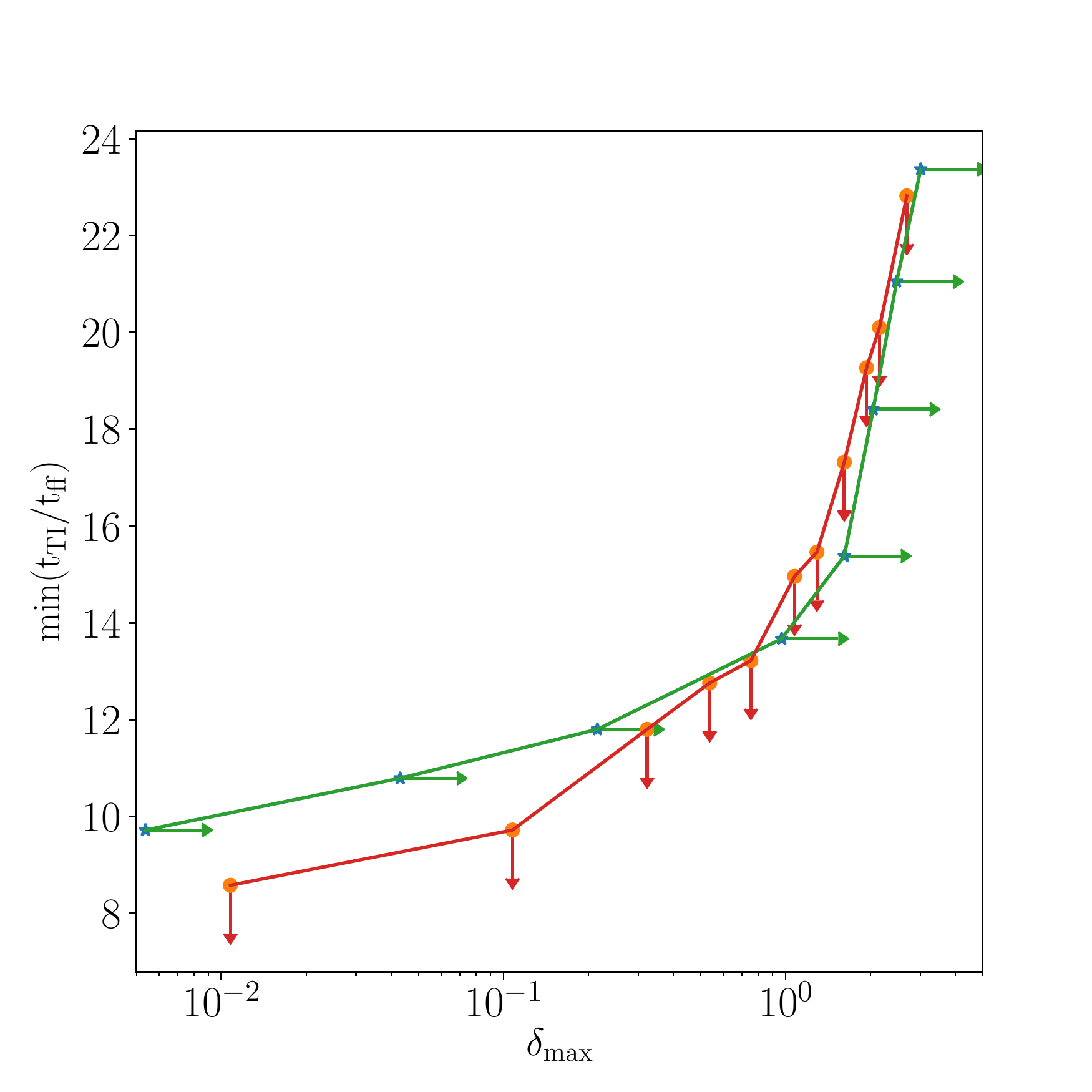}
 \caption{The parameter space for cold gas formation with NFW gravity showing the maximum value of min$(t_{\rm TI}/t_{\rm ff})$ that shows cold gas condensation for a given amplitude of perturbation (denoted by vertical down arrows) and the minimum amplitude of density perturbations required for a given background profile and min$(t_{\rm TI}/t_{\rm ff})$ (denoted by horizontal right arrows). The two curves may not be identical as in one case the background is changing while in another the perturbations in the background are changing in amplitude.}\label{fig:fig3}
\end{figure}
In this section we build up our condensation curve based on the NFW-only simulations. We perform a series of runs with
different amplitudes of initial perturbations for different average $t_{\rm cool}/t_{\rm ff}$ profiles that are set by varying $K_0$ (see section \ref{sec:eq}). Figure \ref{fig:fig2}
shows the 1D profile of shell-averaged number density and $t_{\rm cool}/t_{\rm ff}$ for the runs with $\delta_{\rm max} = 0.32$. We have also shown the corresponding root mean square $\delta$ (Eq. \ref{eq:eqdelta}) and $\delta_{\rm max}$ as a function of radius .

\begin{figure*}
\includegraphics[width=\textwidth]{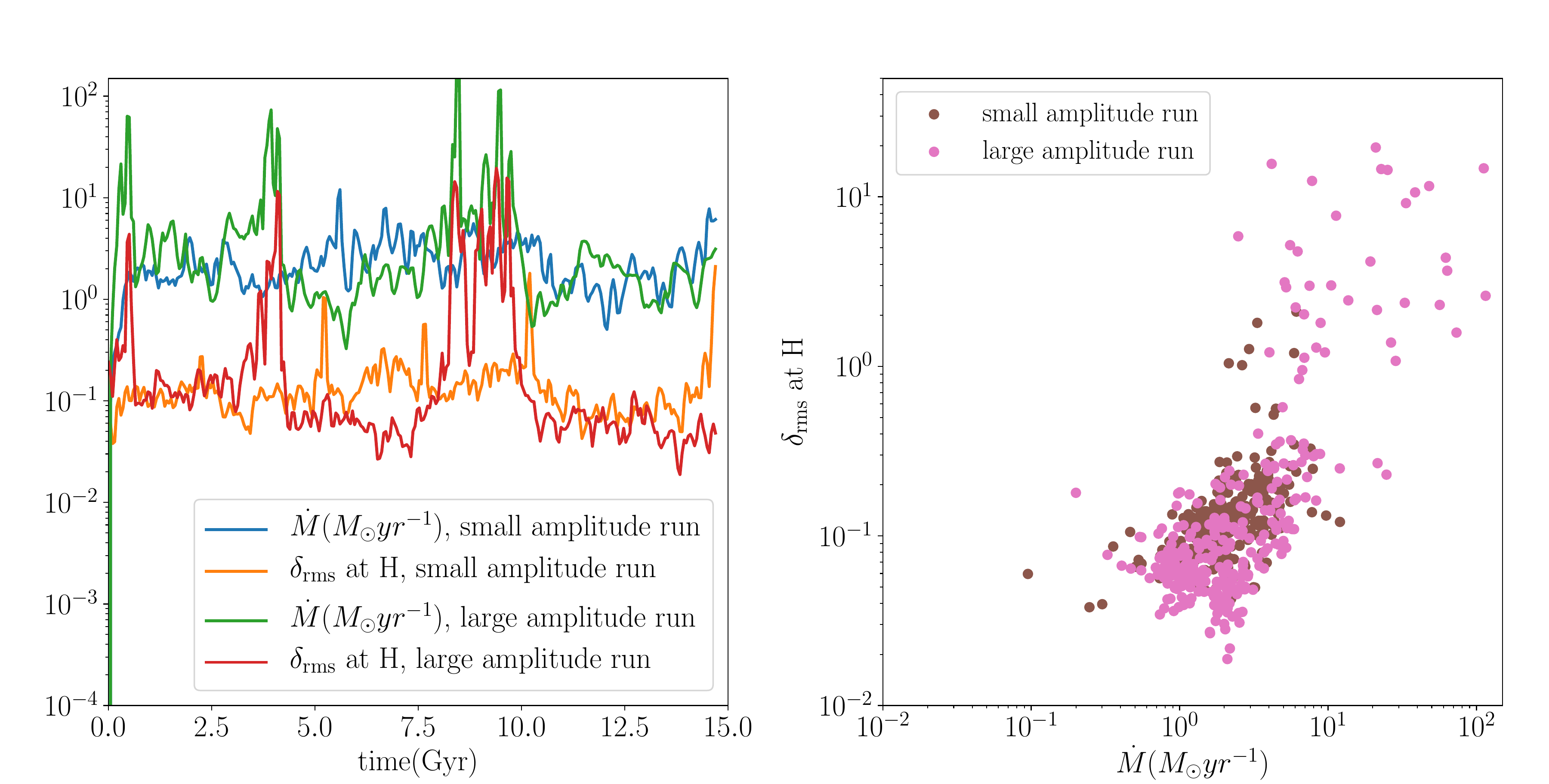}
 \caption{The relation between the mass accretion rate at the inner boundary ($\dot{M}$) and the rms density perturbation amplitude ($\delta_{\rm rms}$) measured at the location ($H$; averaged between $0.9H$ and $1.1H$) of min$(t_{\rm TI}/t_{\rm ff})$ for the low ($\delta_{\rm max} = 0.32$) and high ($\delta_{\rm max} = 1.6$) amplitude runs with $K_0 = 8~{\rm keVcm}^{2}$ $\&$ $K_{100} = 80~{\rm keVcm}^{2}$. Left panel: evolution of $\dot{M}$ and $\delta_{\rm rms}$ with time. Right panel: the correlation between $\dot{M}$ and $\delta_{\rm rms}$ at $H$.}\label{fig:fig4}
\end{figure*}
\subsubsection{Defining the parameter space}
Figure \ref{fig:fig3} shows the condensation curve, to the right/bottom of which there is cold gas formation due to local thermal instability 
(\ref{sec:nfw_lowhigh}). We have a set of runs in which the amplitude of perturbations ($\delta_{\rm max}$) is fixed
and we try to investigate up to what maximum value of min$(t_{\rm TI}/t_{\rm ff})$ at ${t=0}$, we can observe the formation of cold gas by $15$ Gyrs. 
The corresponding curve has vertical down arrows showing the region which forms multiphase gas. We have another set in which we fix a 
background profile with a given value of min$(t_{\rm TI}/t_{\rm ff})$ for each run and we find out the minimum amplitude of density perturbation which triggers the formation of cold gas. The horizontal right arrows signify the condensation zone. 
This plot shows that given a large enough amplitude, the background profile with even quite high min$(t_{\rm TI}/t_{\rm ff})$, can form cold 
gas. 
\subsubsection{Correlation of $\dot{M}$ at the inner boundary and $\delta_{\rm rms}$ at min$(t_{\rm TI}/t_{\rm ff})$}

The crucial question to explore, given a background min$(t_{\rm TI}/t_{\rm ff})$, is how the core with large perturbations differs from
a core with small perturbations. The onset of condensation is expected to be seen within the location of min$(t_{\rm cool}/t_{\rm ff})$ first and at least a fraction of the cold gas thus formed, is expected to move inward. So we expect the mass accretion rate through the inner boundary to shoot up as the gas condenses out.
In this section, we investigate how $\delta_{\rm rms}$ at min$(t_{\rm cool}/t_{\rm ff})$ and $\dot{M}$ measured at the inner boundary are correlated.

The left panel in Figure \ref{fig:fig4} shows the correlation between accretion across the inner boundary and the density fluctuation ($\delta_{\rm rms}$) at the position of min$(t_{\rm TI}/t_{\rm ff})$. The red and green solid lines belong to a run with large initial perturbations ($\delta_{\rm max} \approx 1.6$) while orange and blue solid lines 
correspond to a run with small initial perturbations ($\delta_{\rm max} \approx 0.32$) throughout the ICM. We calculate $\dot{M}$ at the inner boundary ($r_1$) as
\ba
\dot{M} = \iint{r_1}^2 \rho_1 v_1 \sin(\theta) d\theta d\phi
\ea
The rms amplitude is calculated in the following way
\ba
\label{eq:eqdelta}
{\delta_{\rm rms}(r)}^2 = \frac{\iint[(\rho(r,\theta,\phi) - \bar{\rho}(r))/\bar{\rho}(r)]^2 r^2 \sin(\theta) d\theta d\phi}{\iint{r^2 \sin(\theta) d\theta d\phi}}
\ea
where $\bar{\rho}(r)$ is shell-averaged density. Now $\delta_{\rm rms}(r)$ is averaged within $0.9 H$ and $1.1H$ where $H$ is the radius of
min$(t_{\rm TI}/t_{\rm ff})$. 

It is evident that accretion is highly correlated with $\delta_{\rm rms}$ particularly for a reasonably large $\delta_{\rm rms}$ at $H$ (see right panel of Figure \ref{fig:fig4}). For smaller perturbations, the density fluctuations growing at min$(t_{\rm TI}/t_{\rm ff})$ sometimes may not reach the inner boundary and remain suspended within the medium or
mix with the hot surroundings before reaching the inner boundary. For the large perturbation amplitude, large cold blobs form and easily decouple from the ICM and hence the correlation is more prominent specifically at large $\delta_{\rm rms}$ (although with a large scatter). 

\subsection{NFW+BCG: the condensation curve}
\label{sec:resnfwbcg}
\begin{figure*}
\includegraphics[width=\textwidth]{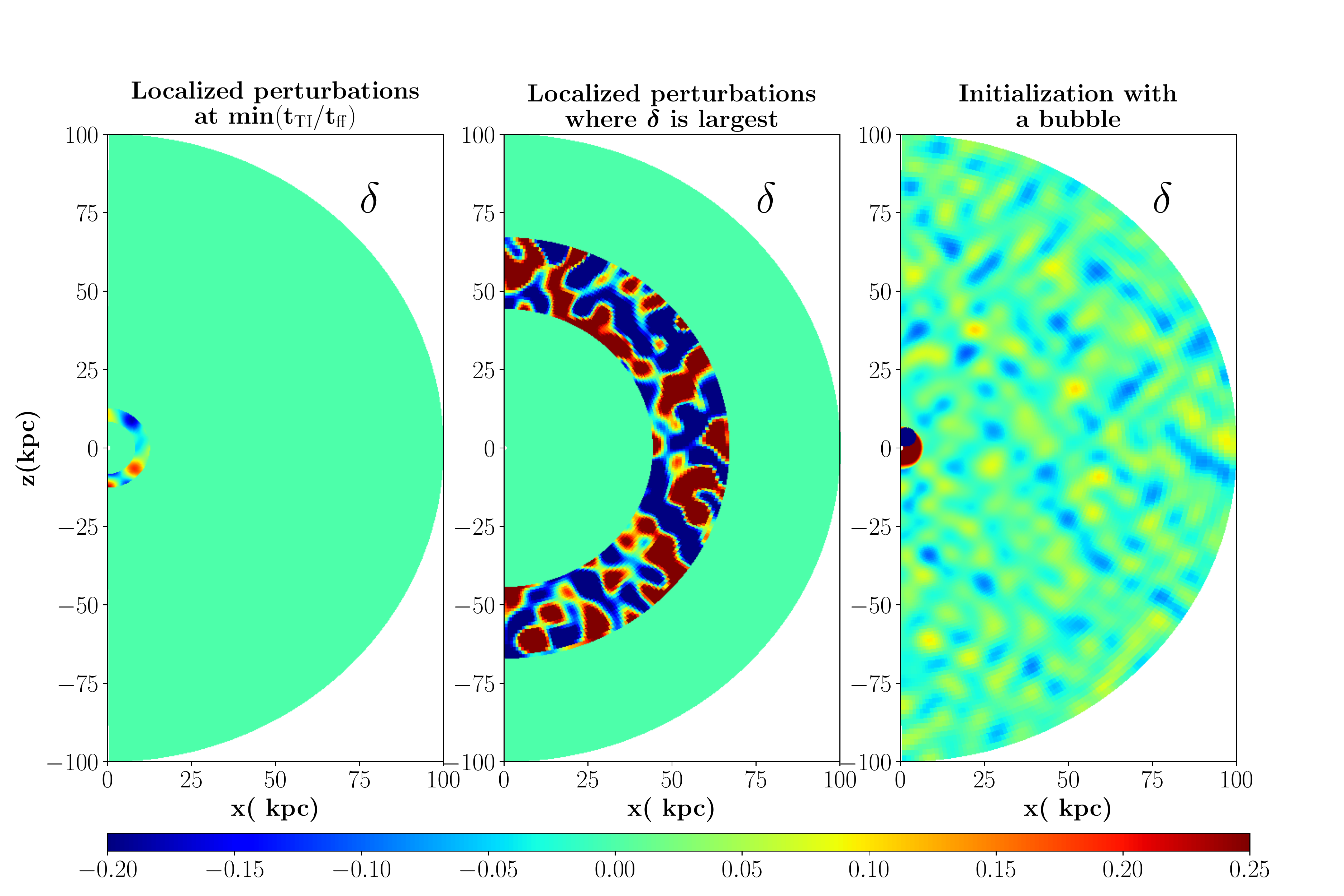}
\caption{The initial overdensity ($\delta$) for the two cases with localized perturbations (left two panels) and a low-density bubble (right panel) in the ICM with the NFW+BCG potential and the background entropy parameter $K_0 = 5~{\rm keVcm}^{2}$ corresponding
 to min$(t_{\rm TI}/t_{\rm ff})= 8.18$. In the third panel, there is apparently a dense core at the center because it shows the 
 density contrast between the background and the bubble.}\label{fig:fig5}
\end{figure*}

\begin{figure*}
\includegraphics[width=\textwidth]{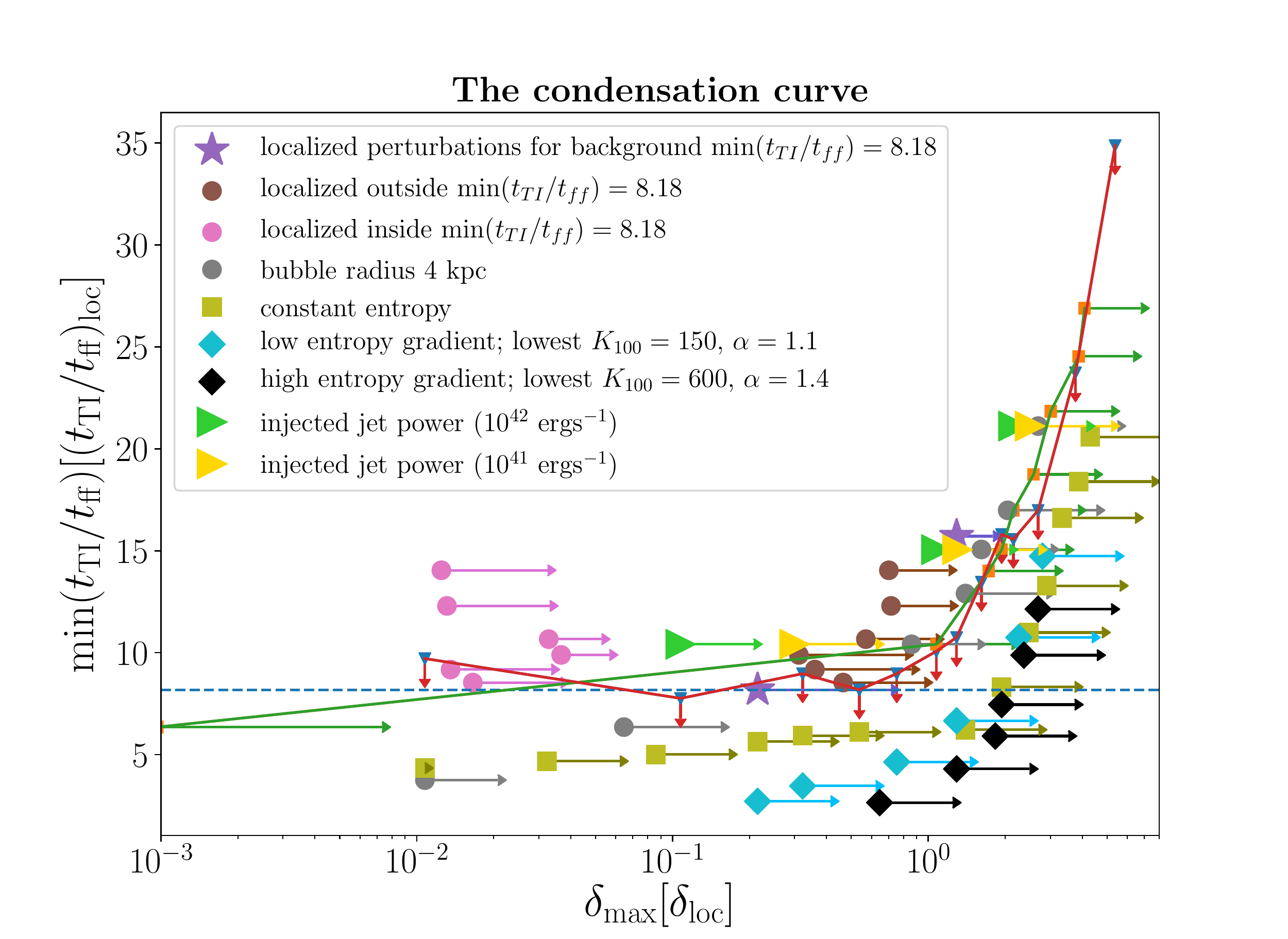}
\caption{The condensation curve similar to Figure \ref{fig:fig3}, but with the NFW+BCG potential and varying different parameters.
The red and green solid lines correspond to the runs with perturbations present throughout the cluster and form the backbone of this figure. 
Below this curve, the CGM is susceptible to condensation. The threshold min$(t_{\rm TI}/t_{\rm ff})$ for multiphase condensation is shifted
significantly for $\delta \lesssim 1$ under different conditions (jet, bubble, entropy variation). Jets enhance condensation while buoyant bubble and steep entropy gradients suppress it. Somewhat counterintuitively, even flat entropy (throughout) systems condense out at somewhat larger amplitudes compared to the fiducial green line. For localized perturbations, there is a small scatter around the condensation curve
because local entropy gradients are different in the central region and in the outskirts. Note that the two parameters ($t_{\rm TI}/t_{\rm ff}$ and $\delta$) in the case of localized runs are calculated locally.
The thin dashed line corresponds to min$(t_{\rm TI}/t_{\rm ff})=8.18$, the background value for the runs with localized perturbations.}\label{fig:fig6}
\end{figure*}
In this section we present the condensation curve for the more realistic case of an NFW+BCG potential
(see Figure \ref{fig:fig6}). The presence of BCG reduces the threshold $\rm{min(t_{TI}/t_{ff})}$ to some extent for lower amplitudes. 
But again, with high enough amplitudes, any background profile seems to eventually form cold gas. The underlying red and green lines are for the cases where perturbations are present throughout the medium (following section \ref{sec:nfwbcg_lowhigh}). The arrows show the parameter that we vary
(min$[t_{\rm TI}/t_{\rm ff}]$ or $\delta_{\rm max}$) until cold gas is seen, while the other parameter is held fixed.

\subsubsection{Localized perturbations}
As we mention at the beginning of the section \ref{sec:result}, the basic condensation curve is built out of two parameters min$(t_{\rm TI}/t_{\rm ff})$ and $\delta_{\rm max}$ where 
the latter denotes the value at the maximum amplitude of the overdensity field.
These parameters only label the two variables throughout the domain. The shell-averaged values of $t_{\rm TI}/t_{\rm ff}$ and 
$\delta$, are different at different radii. We wish to study the condensation curve for localized perturbation with a given value of
$\delta$ and $t_{\rm TI}/t_{\rm ff}$.

We consider localized perturbations following
section \ref{sec:l_nfwbcg}. The two `stars' in Figure \ref{fig:fig6} represent the points corresponding to 
the two cases where perturbations are present only at the radius of min$(t_{\rm TI}/t_{\rm ff})$ and the radius where our homogeneous perturbation field 
has the maximum amplitude ( $\approx 55$ kpc with background $t_{\rm TI}/t_{\rm ff} \approx 15$). We increase $\delta$ until condensation happens. For these
cases of localized perturbations, $(t_{\rm TI}/t_{\rm ff})$ and $\delta$ are calculated around the radius ($R_1$) of maximum $\delta = \delta_{\rm max}$, within $0.99R_1$ and $1.01R_1$. 
The `stars' follow a similar trend (with $10\%$-$20\%$ variation in the threshold $t_{\rm TI}/t_{\rm ff}$). Relatively, with localized perturbations it is slightly more difficult at the plane of min$(t_{\rm cool}/t_{\rm ff})$ but slightly easier at the plane of $\delta_{\rm max}$
to condense than with perturbations throughout. But the general behavior of the condensation curve with the changes in $t_{\rm TI}/t_{\rm ff}$ and $\delta$ (either min$(t_{\rm cool}/t_{\rm ff})$ and $\delta_{\rm max}$ or $[t_{\rm TI}/t_{\rm ff}]_{\rm loc}$ and $\delta_{\rm loc}$)
remains similar.  

It is also interesting to explore the condensation curve for localized perturbations seeded in the shells outside and inside the location of min$(t_{\rm cool}/t_{\rm ff})$. 
The brown and pink circles denote these runs respectively. If we put these points 
according to the background min$(t_{\rm cool}/t_{\rm ff})=8.18$, they will fall on the horizontal line $y=8.18$. This line will cut the condensation curve at around $\delta \lesssim 0.1$.
All the pink points correspond to amplitudes less than that while all the brown points correspond to greater amplitude. This implies it is harder to form cold gas outside the 
radius of min$(t_{\rm cool}/t_{\rm ff})$ rather than inside it.
This can also be understood in terms of the entropy profile which is shallower at smaller radii (section \ref{sec:cases}). 

Note that the condensation 
curve for localized perturbations (marked by the local values of the parameters and shown by the pink and brown points and the `stars' in Figure \ref{fig:fig6}) will
deviate from the solid green line, which corresponds to the perturbations seeded everywhere. Figure 8 in \citealt{ashmeet14} also shows difference in condensation of a blob
inside and outside min$(t_{\rm cool}/t_{\rm ff})$.
\subsubsection{Effects of a bubble and outflows}
\label{sec:cases}
In this section, first we plot the points corresponding to runs with perturbations throughout the domain but with
a low-density bubble near the inner boundary (following section \ref{sec:bnfwbcg}). Placing the
bubble near the center, we fix the background $t_{\rm TI}/t_{\rm ff}$ while varying $\delta_{\rm max}$ to find the threshold amplitude
for condensation. In Figure \ref{fig:fig6} these are marked by grey points with right grey arrows. The presence of a bubble slightly enhances the
$t_{\rm TI}/t_{\rm ff}$ due to low density. The effect of the bubble is more noticeable for smaller $t_{\rm TI}/t_{\rm ff}$ because the bubble mixes with the background ICM and raises the effective cooling time before cold gas condenses. So a bubble actually prevents cold gas formation as it buoyantly rises. When
min$(t_{\rm TI}/t_{\rm ff})$ is high and there are large density perturbations anyway, the presence of a bubble does not make a difference and the trend follows
the original condensation curve. Bigger/Smaller bubbles do not change the trend significantly.

We introduce jets of constant power into an identical ICM setup (following section \ref{sec:jnfwbcg}). The jets enhance cold gas formation as we observe condensation for smaller perturbation amplitudes. For two different jet powers $\approx 10^{41}~{\rm ergs}^{-1}$ and $\approx 10^{42}~{\rm ergs}^{-1}$ (varying $\dot{M}_{\rm acc}$) we have three runs
each, corresponding to different initial min$(t_{\rm TI}/t_{\rm ff})$ (given by lime-green and golden right-triangles in Figure \ref{fig:fig6})
where we see that with higher jet power it is mildly easier for cold gas to condense. Precisely, for the lowest min$(t_{\rm TI}/t_{\rm ff})$, the threshold $\delta$ required is $3$ 
times higher than that with higher jet power. Our jet injection increases the local density contrast of the CGM, rather than heat it, thereby making multiphase condensation easier.


\subsection{NFW+BCG: background entropy variation}

For the next three cases (light green, blue and black squares in Figure \ref{fig:fig6}), we initialize the ICM
with a constant entropy and a radially varying power-law entropy respectively (section \ref{sec:entrv}). We try the second case for two different entropy gradients (blue: lower gradient, $\alpha=1.1$; black: higher gradient, $\alpha=1.4$). For the first case, the threshold condensation curve is slightly lower compared to the fiducial curve which has both an entropy core and a power-law in the outskirts. This implies that it is more difficult to condense in this regime. However, with constant background entropy we also see that once condensation is triggered, there is a large amount of gas condensing out of the hot medium. We discuss this further in the next section. The runs with large positive entropy gradients form cold gas at a much higher threshold amplitude, which is raised a little more for a larger gradient. Beyond $\delta \gtrsim 1$
the condensation curve converges for all the cases and provides substantial evidence that once the ICM has a very large density perturbation, it will start forming cold gas essentially irrespective of the background $t_{\rm TI}/t_{\rm ff}$.

\begin{table*}
 \caption{Numerical experiments to quantify the amount and extent of cold gas for different entropy profiles. We use the following parameters for all the runs: $\delta_{\rm max}=0.2$, $n_{e,{\rm out}} = 0.035~{\rm cm}^{-3}$ and each run corresponds to $0.65q_{\rm thresh}$}
 
{\centering
\begin{tabular}{c c c c c c}
\hline\hline
 Entropy &min$(t_{\rm TI}/t_{\rm ff})$ &min$(t_{\rm TI}/t_{\rm ff})$ & Location of min$(t_{\rm TI}/t_{\rm ff})$ &Average cold mass ($M_{\odot}$)& $r_{\rm cold, max}$ (kpc)\\
 &($q_{\rm thresh}$; compare with Figure \ref{fig:fig6})&($0.65q_{\rm thresh}$)&(in kpc)&&\\
 &at $\delta_{\rm max}=0.2$&&&&\\

\hline
 $K_0K_{100}$ & $9.8$&$6.4$ &$17.3$&$2.55\times10^{8}$&$15.26$ \\
 $K_0$&$6.0$&$3.9$&$99.09$&$1.75\times10^{9}$&$75.66$\\
 $K_{100}$&$3.88$&$2.5$&$1.009$&$1.56\times10^{7}$&$10.84$\\
 \hline
\end{tabular}
} \\
\textbf{Notes:} $r_{\rm max}$ is the maximum radius upto which cold gas is obtained for all times. Note that $q_{\rm thresh}$ for each of these cases fall almost on top of the condensation curves respectively, in Figure \ref{fig:fig6}, which implies $n_{e,{\rm out}}$ does not affect the curve.
 \label{table:listofruns2}
\end{table*}

\subsubsection{Amount of cold gas with entropy variation: implications for the CGM}
\begin{figure}
\includegraphics[width=0.5\textwidth]{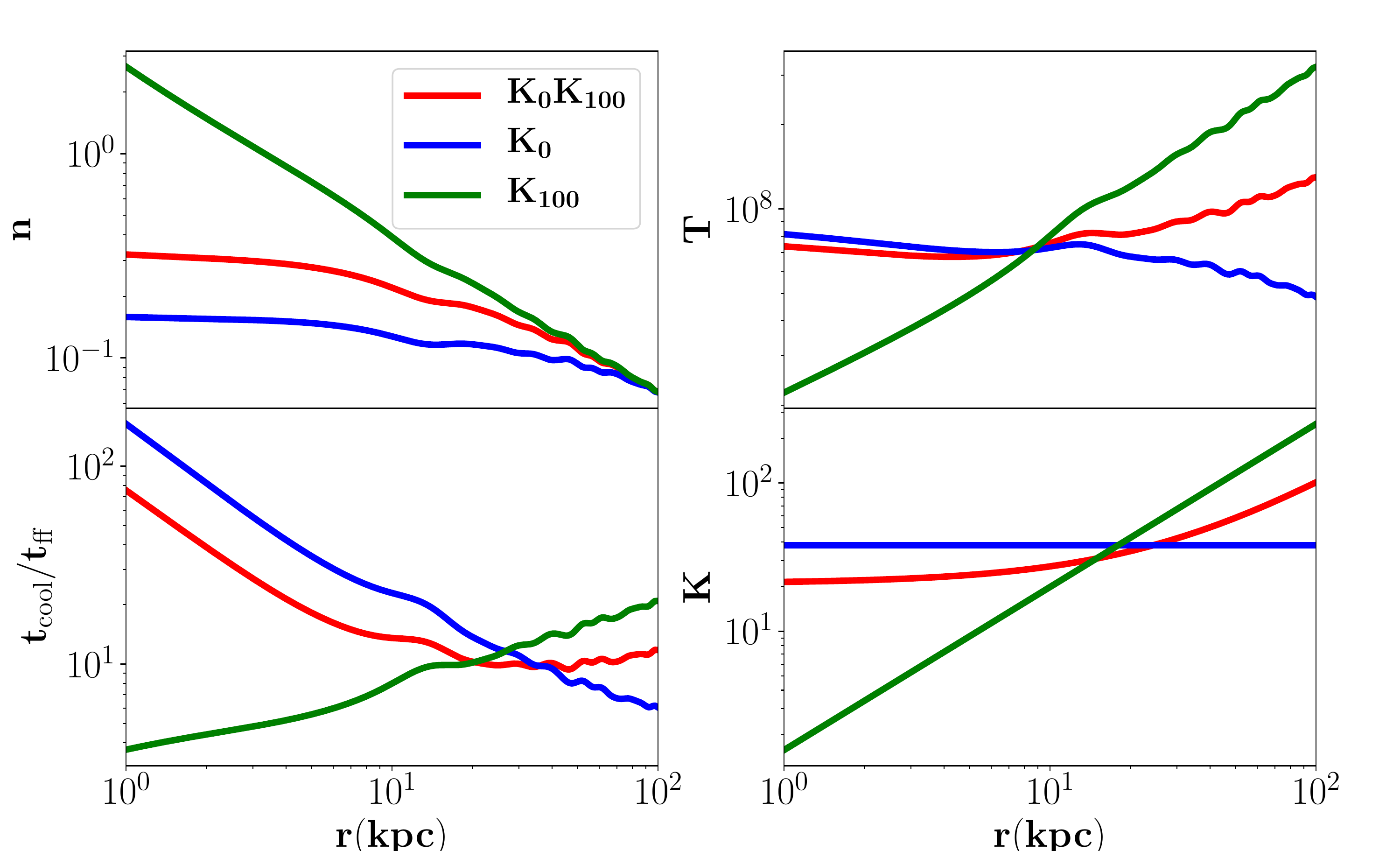}
\caption{The radial profiles for runs with different entropy profiles ($K_0K_{100}$, $K_0$ and $K_{100}$) at the threshold for condensatio ($q_{\rm thresh} = {\rm min}[t_{\rm TI}/t_{\rm ff}]$) for $\delta_{\rm max}=0.2$. These three cases are considered for the comparisons we do in section \ref{sec:compentr}. All these runs have $n_{e,{\rm out}}= 0.035~{\rm cm}^{-3}$. Although density profiles are very different, the total gas mass (which is dominated by the largest radii) is very similar for the three cases.}
\label{fig:fig9}
\end{figure}

\label{sec:compentr}
Because of numerical problems at the outer boundary with a realistic density, we have runs with constant entropy throughout in which we have increased $n_{\rm e, out}$ (section \ref{sec:ini_bound}). We need to assure that the condensation curve is not affected due to a different outer density. Hence we test cases with different entropy profiles but using this new outer density. We compare three cases with different profiles: one with a constant inner entropy and outer power-law ($K_0K_{100}$), one with a constant entropy throughout the medium ($K_0$), and one with power-law entropy ($\alpha=1.1$, see section \ref{sec:eq}) throughout the medium ($K_{100}$). For all the runs we keep the outer boundary at $n_{e,{\rm out}} = 0.035~{\rm cm}^{-3}$. We fix the maximum amplitude $\delta_{\rm max} = 0.2$ and scan across a range of background min$(t_{\rm TI}/t_{\rm ff})$ to see up to what threshold value of this parameter (second column in Table \ref{table:listofruns2}) we see condensation. The radial profiles for the backgrounds with the threshold $t_{\rm TI}/t_{\rm ff}$ that just show cold gas, are shown in Figure \ref{fig:fig9}. The corresponding points for $K_0K_{100}$ and $K_{100}$ coincide with the ones shown in the respective condensation curves (match the values in the second column of Table \ref{table:listofruns2} with Figure \ref{fig:fig6} for $\delta_{\rm max} = 0.2$). This way we verify that the outer density does not affect the condensation curve. 

\begin{figure}
\includegraphics[width=0.5\textwidth]{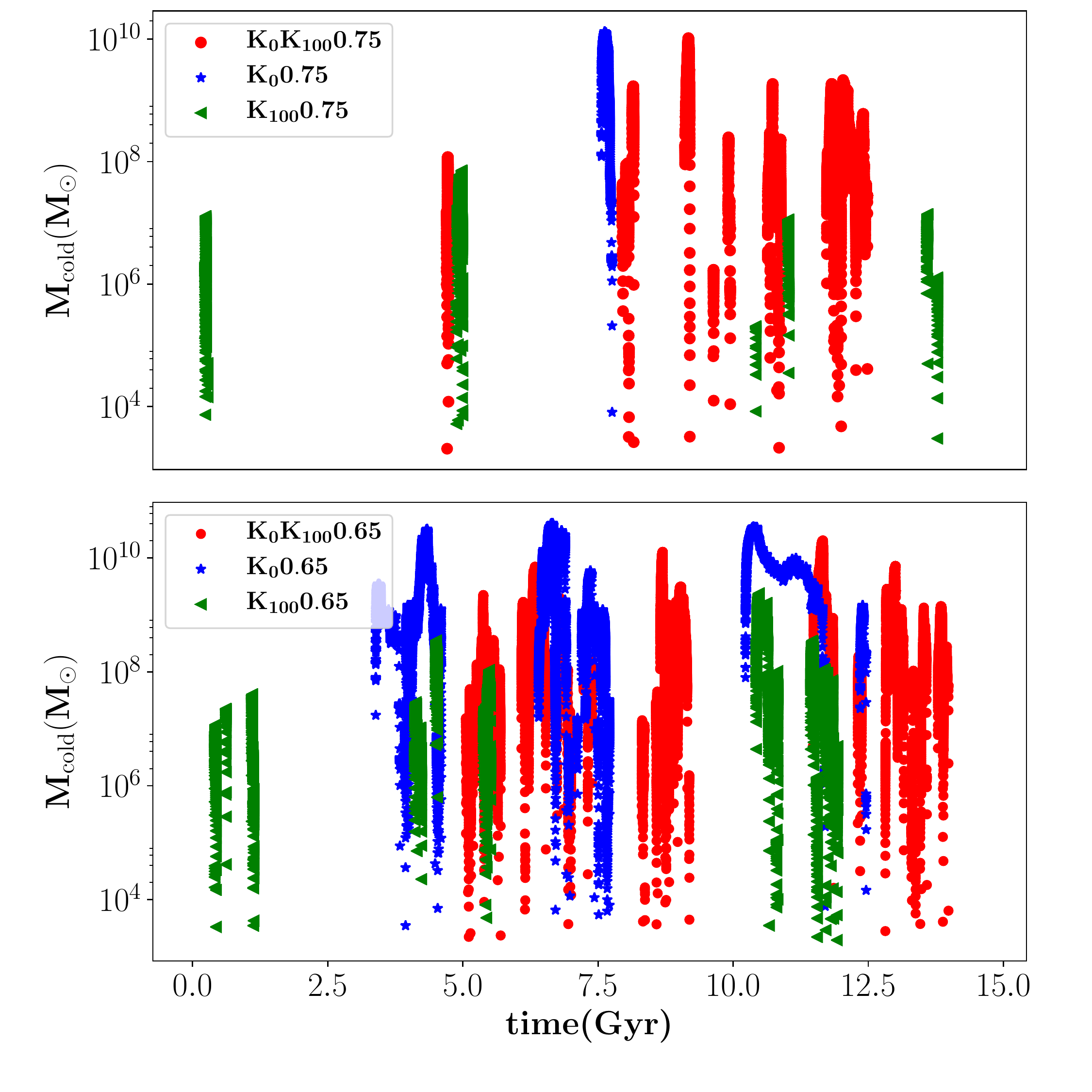}
\caption{The total cold gas mass as a function of time for different entropy profiles ($K_0K_{100}$, $K_0$ and $K_{100}$) with $0.75 q_{\rm thresh}$ and $0.65 q_{\rm thresh}$ where $q = {\rm min}(t_{\rm TI}/t_{\rm ff})$. Total gas condensing out increases as we move into the condensation zone ($0.75 q_{\rm thresh}$ to $0.65 q_{\rm thresh}$). Additionally, for $K_0$ the total amount of cold gas grows faster than for $K_0K_{100}$ and $K_{100}$, which further implies that larger cores give rise to enhanced condensation inside the condensation zone. }\label{fig:fig7}
\end{figure}

\begin{figure}
\includegraphics[width=0.5\textwidth]{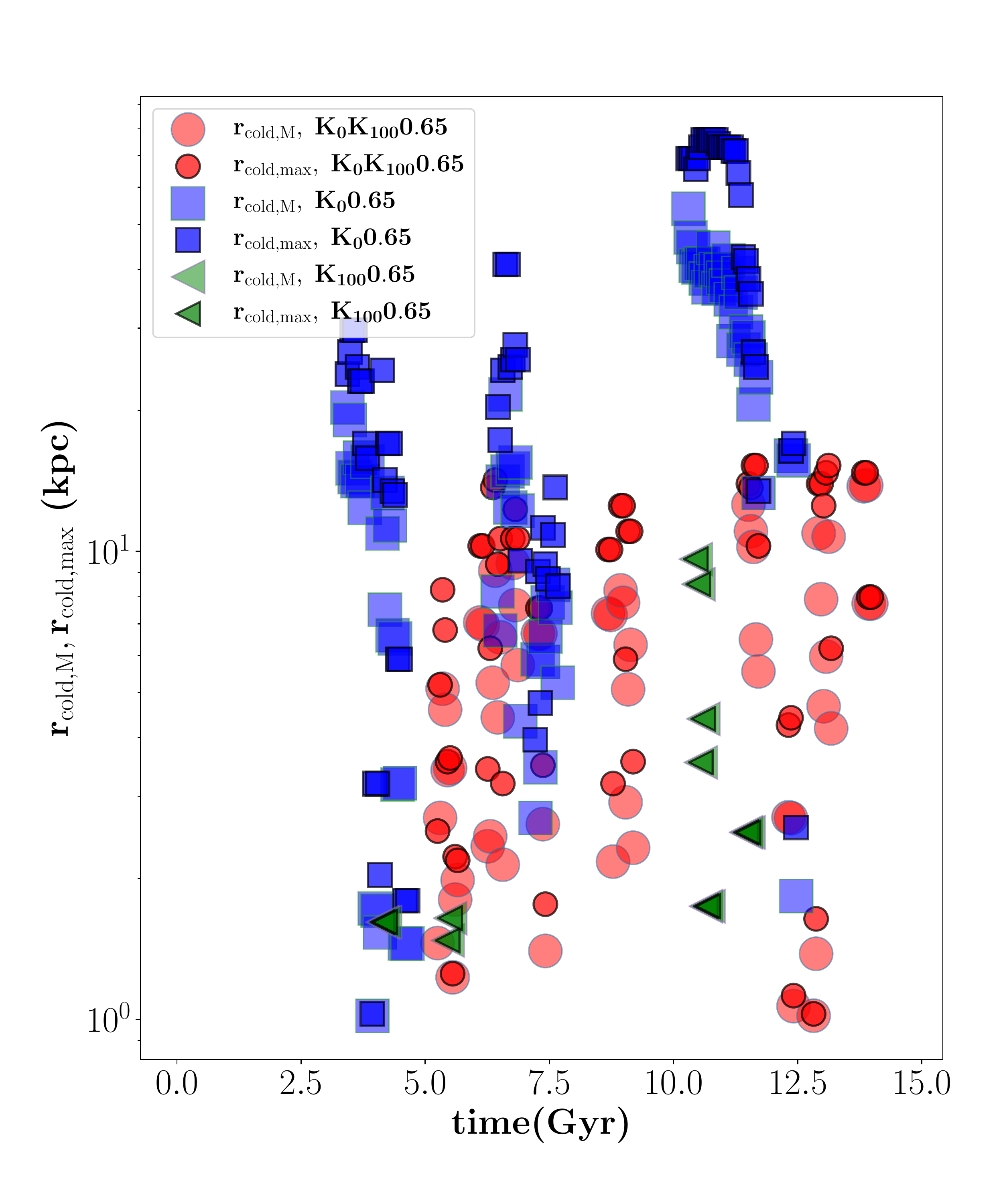}
\caption{The radius of cold gas ($r_{\rm cold}$) as function of time for the three entropy profiles $K_0K_{100}$, $K_0$ and $K_{100}$ with $0.65 q_{\rm thresh}$, where $q = {\rm min}(t_{\rm TI}/t_{\rm ff})$. The bigger circles/squares/triangles represent the mass averaged radius of the cold gas ($r_{\rm cold, M}$; eq \ref{eq:rcold}) and the circles/squares/triangles with black border correspond to the maximum radius where cold gas is present ($r_{\rm cold, max}$). Note that in $K_{100}$, $r_{\rm cold, M}$ and $r_{\rm cold, max}$ are coincident. }\label{fig:fig10}
\end{figure}

We also consider these three cases to compare the amount of cold gas produced in each case. For that, we do a few numerical experiments keeping the same amplitude of perturbation ($\delta_{\rm max} = 0.2$) and lower the $q = {\rm min}(t_{\rm TI}/t_{\rm ff})$ below the threshold ($q_{\rm thresh}$), that is, we move into the condensation zone. We try three different entropy profiles with $0.75 q_{\rm thresh}$, $0.7 q_{\rm thresh}$ and $0.65 q_{\rm thresh}$ respectively. Figure \ref{fig:fig7} shows the total cold gas mass as a function of time in the runs with $0.75 q_{\rm thresh}$ and $0.65q_{\rm thresh}$. As expected we see an increase in the cold gas mass as we move into the condensation zone. The cases with large cores give maximum cold gas inside the condensation zone. The cases with $0.7q_{\rm thresh}$ follow a similar trend. The maximum radial extent up to which cold gas is seen in $K_0$ is also the largest as shown in Table \ref{table:listofruns2}, which also gives the amount of cold gas averaged over the entire time. This implies that for the CGM (which have larger cores), thermal instability models predict a higher amount of cold gas spread over larger radii. Figure \ref{fig:fig10} shows the mass-weighted cold gas radius ($r_{\rm cold}$) and the maximum cold gas radius as functions of time. We calculate mass-weighted $r_{\rm cold}$ as:
\ba
\label{eq:rcold}
 r_{\rm cold, M} = \int r_{\rm cold} dM_{\rm cold}/\int dM_{\rm cold}
 \ea
Figure \ref{fig:fig8} shows the gas mass fraction averaged over time for backgrounds with different fractions of $q_{\rm thresh}$. The amount of cold gas in the run with entropy gradient is much smaller compared to the other two (flat entropy, core+power-law entropy) cases. Not only is the mass of the cold gas lower, but the rate of increase of the cold gas mass with a decrease in $q$ (a fraction of $q_{\rm thresh}$) is also smaller for this case.


\section{Discussion and Astrophysical implications}
\label{sec:disc}
\subsection{Significance of the condensation curve}
\label{sec:condcrv}
\par{The condensation curve defines a zone of condensation in an idealized parameter space corresponding to any CGM environment. We do not quantify in detail how much gas condenses out (except in Figure \ref{fig:fig7} and Figure \ref{fig:fig8}) within the condensation zone and whether it is sufficient for star formation. Instead we simply delineate the parameter space
for which multiphase condensation occurs due to thermal instability. It is likely that more cold gas will be formed away from the 
condensation curve into the zone of condensation (Figure \ref{fig:fig7}). For $\delta_{\rm max} \lesssim 1$ ($\delta_{{\rm min}(t_{\rm cool}/t_{\rm ff})} < \delta_{\rm max}$), valid for cool core clusters, the threshold value of background min$(t_{\rm cool}/t_{\rm ff})$ plays a pivotal role in determining whether multiphase condensation is possible.
This corroborates earlier works which consider small, isobaric perturbations. The interesting feature of the condensation curve is that it steeply rises beyond the amplitude of $\delta_{\rm max} \sim 1$. This makes it evident that high amplitude perturbations trigger condensation, almost irrespective of the ratio of the background $t_{\rm cool}/t_{\rm ff}$. The relevance of a threshold min$(t_{\rm cool}/t_{\rm ff})$ of the background medium loses significance in this regime. 
This may explain the cold gas formation for the backgrounds with min$(t_{\rm cool}/t_{\rm ff})$ ratio as high as $15-30$ (\citealt{hogan17}). However, note that cold gas can last longer if it has angular momentum and the ICM can be seen in a state of min$(t_{\rm cool}/t_{\rm ff}) \gtrsim 10$ even if the cold gas originated in a state with the core lying below the threshold min$(t_{\rm cool}/t_{\rm ff}) \lesssim 10$ (\citealt{prasad15}, \citealt{prasad18}).

\begin{figure}
\includegraphics[width=0.5\textwidth]{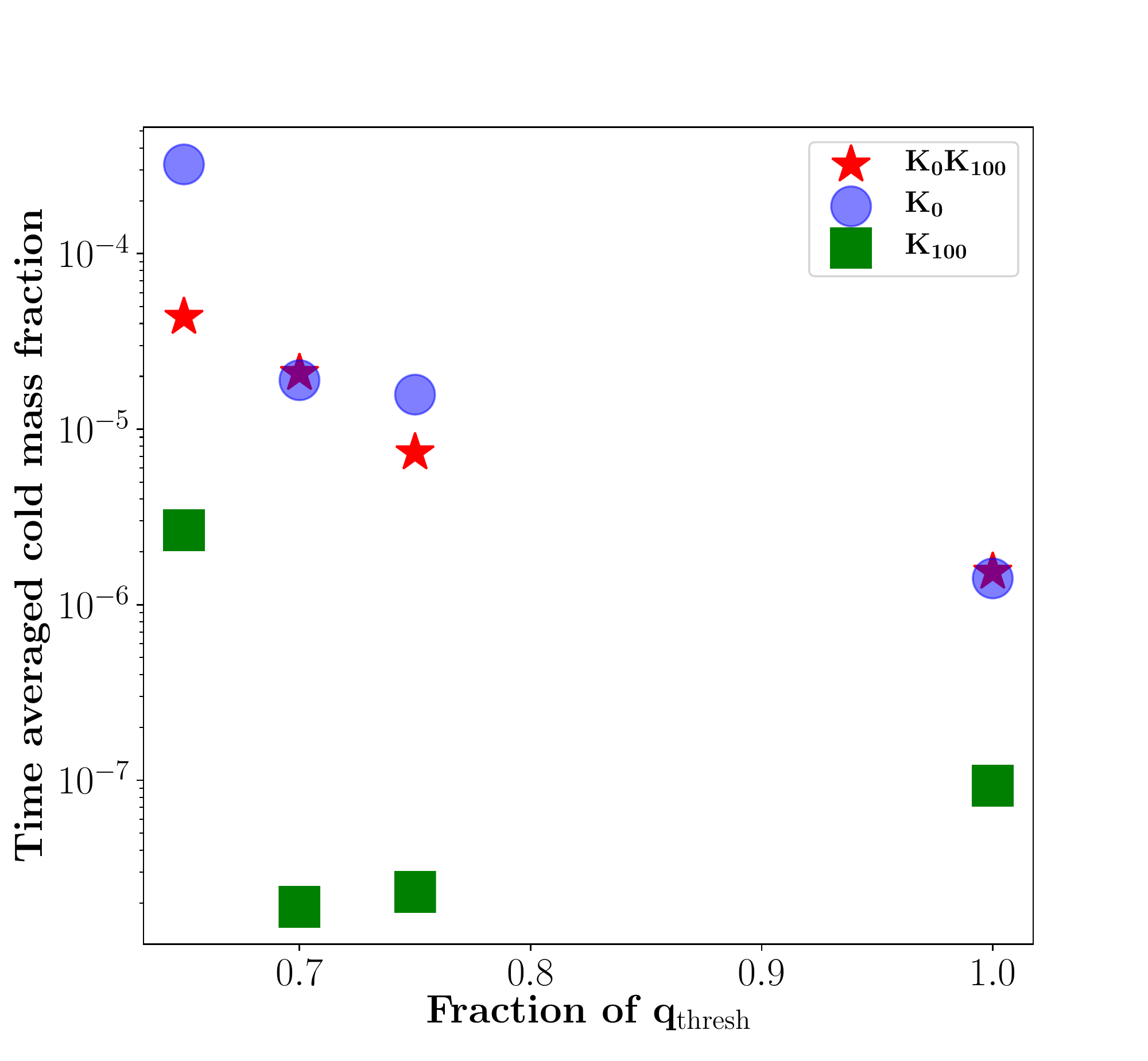}
\caption{The time-averaged (over the entire 15 Gyr) cold gas mass for different entropy profiles ($K_0K_{100}$, $K_0$ and $K_{100}$) as a function of $q_{\rm thresh}$, $0.75 q_{\rm thresh}$, $0.7 q_{\rm thresh}$ and $0.65 q_{\rm thresh}$ where $q = {\rm min}(t_{\rm TI}/t_{\rm ff})$. This shows that on an average the run $K_0$ starts condensing larger amount of cold gas and this is also true deeper into the condensation zone. Note that the amount of cold gas for the power-law entropy profile ($K_{100}$) at $0.65q_{\rm thresh}$ is comparable to the other cases ($K_0$, $K_0K_{100}$) at $q_{\rm thresh} = 1$ and the rise in the mass fraction is maximum for a flat entropy profile.}\label{fig:fig8}
\end{figure}

The location of the large density contrast is also important for the susceptibility to condensation. From our localized perturbation runs it is evident that
inside the region of min$(t_{\rm cool}/t_{\rm ff})$, close to the center, condensation occurs even for locally high values of
$t_{\rm cool}/t_{\rm ff}$ (of course min$(t_{\rm cool}/t_{\rm ff})$ is still smaller than the threshold). The ease of condensation closer to the center can be interpreted as a consequence of the shallower entropy gradient. 
Interestingly, for a constant entropy throughout the medium, the threshold min$(t_{\rm TI}/t_{\rm ff})$ is smaller than that in core+power-law entropy profile, however, the quantity of gas that condenses out is significantly large and distributed over large radii. }
\begin{figure}
\includegraphics[width=0.5\textwidth]{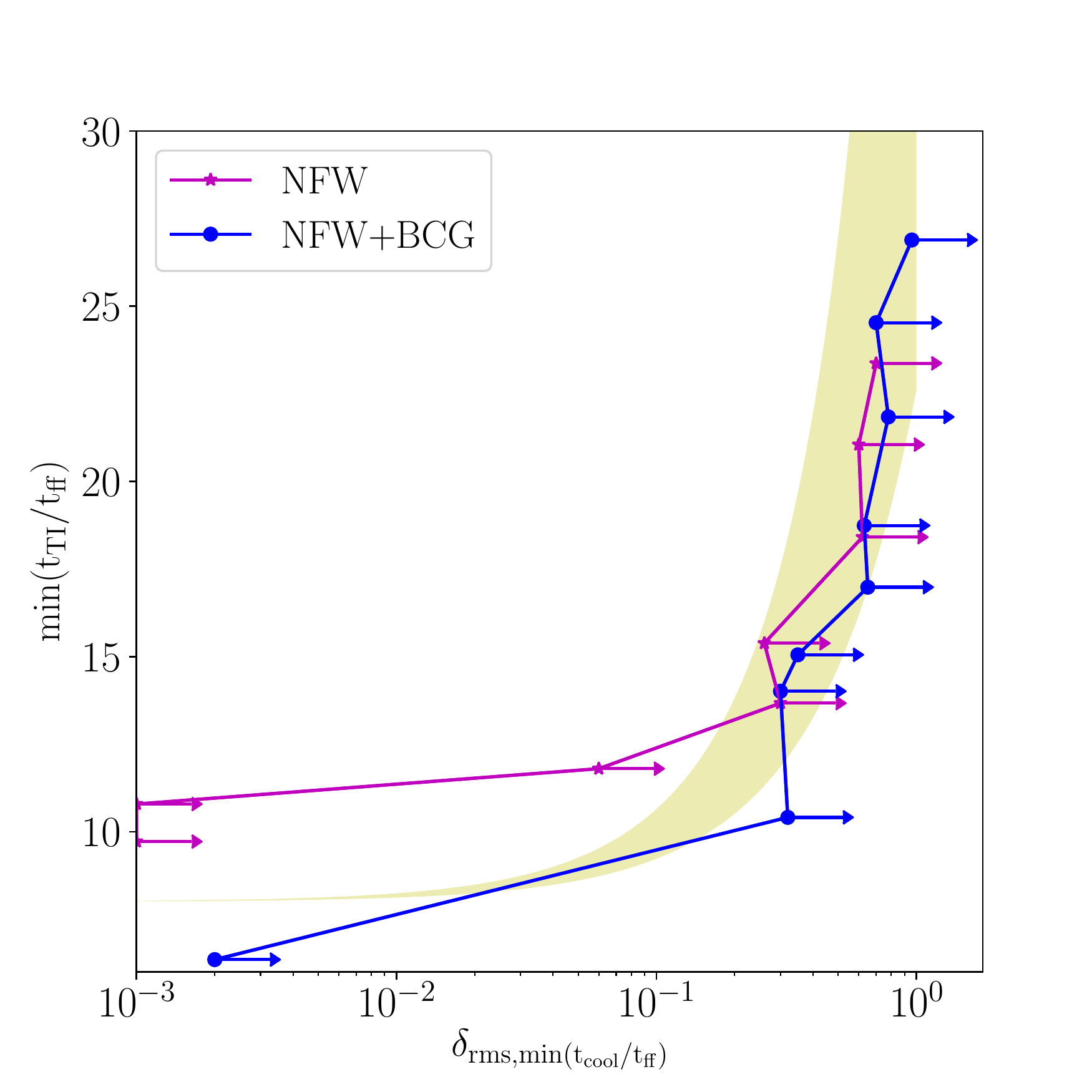}
\caption{ The condensation curve using the average $\delta_{\rm rms}(r)$ ($\lesssim$ local $\delta_{\rm max}$; see red lines in Figure \ref{fig:fig2}) between $0.9H$ and $1.1H$ where $H$ is the location of min$(t_{\rm cool}/t_{\rm ff})$ in each run. The region shaded in yellow corresponds to $t_{\rm cool}/t_{\rm ff}$ of the blob between $\alpha_c=-1$ and $\alpha_c=0.5$ as described in eq. \ref{eq:blob} which provides a good description of the condensation curve derived from our simulations.}\label{fig:fig11}
\end{figure}

We plot a modified version of the condensation curve in Figure \ref{fig:fig11} in which we take the condensation curves for NFW and NFW+BCG and label the perturbation amplitude by $\delta_{\rm rms}(r)$ averaged around the radius ($H$) of min$(t_{\rm cool}/t_{\rm ff})$ between $0.9H$ and $1.1H$. 
We do a simple analytic estimate to understand the condensation curve. For a single blob moving through the medium, we know that the ratio of the cooling time to the free-fall time in a background of density $n_0$, temperature $T_0$, $t_{\rm cool}/t_{\rm ff}={\Big(t_{\rm cool}/t_{\rm ff}\Big)}_0$ and density contrast $\delta$ between the blob and the background, can be expressed as:
\ba
\nonumber
{\Big(\frac{t_{\rm cool}}{t_{\rm ff}}\Big)}_{\rm blob} &=& \frac{\frac{3}{2}n_{\rm blob} k_B T_{\rm blob}}{n^2_{\rm blob}\Lambda(T_{\rm blob})t_{\rm ff, blob}} \\
                   \nonumber
                   &=& \Big(\frac{3}{2}\frac{k_B}{n_{\rm blob}T_{\rm blob}}\Big) \frac{T^2_{\rm blob}}{\Lambda(T_{\rm blob})t_{\rm ff, blob}}.
\ea
Now we consider that the blob is in pressure equilibrium with the medium and hence $n_0T_0 = n_{\rm blob}T_{\rm blob}$ (this is justified because sound-crossing time is short). Therefore, we can simply write $T_{\rm blob} = T_0/(1+\delta)$.  At a given radial location, the free-fall times for the blob and the background are the same: $t_{\rm ff, blob} = t_{\rm ff, 0}$. We also consider a simple form of the cooling function as $\Lambda (T) = \Lambda_0 T^{\alpha_c}$ and obtain the following expression,
\ba
\nonumber
{\Big(\frac{t_{\rm cool}}{t_{\rm ff}}\Big)}_{\rm blob} &=&\Big(\frac{3}{2}\frac{k_B}{n_{0}T_0}\Big) \frac{T^2_0/{(1+\delta)}^2}{\Lambda_0 T^{\alpha_c}_0/{(1+\delta)}^{\alpha_c}t_{\rm ff, 0}}\\
\label{eq:blob}
                         &=& {\Big(t_{\rm cool}/t_{\rm ff}\Big)}_0{(1+\delta)}^{(\alpha_c-2)}.
\ea
The yellow shaded region in Figure Figure \ref{fig:fig11} shows the parameter space corresponding to ${(t_{\rm cool}/t_{\rm ff})}_{\rm blob}=8$ and the varying parameters being ${(t_{\rm cool}/t_{\rm ff})}_0$ and $\delta$ between the values of $\alpha_c=-1$ (crudely mimicking galaxy cooling function) and $\alpha_c=0.5$ (cluster cooling function). This simply shows that the condensation curve traces out the locus of ${\Big(\frac{t_{\rm cool}}{t_{\rm ff}}\Big)}_{\rm blob} =8$, i.e, multiphase condensation in the plane of background min$(t_{\rm cool}/t_{\rm ff})$ happens such that the perturbations (rather than the background) maintain a threshold $(t_{\rm cool}/t_{\rm ff})$. 
Hence we generalize the idea that the background min$(t_{\rm cool}/t_{\rm ff})$ must fall below a threshold for multiphase condensation, to the $(t_{\rm cool}/t_{\rm ff})$ of the overdense region. 

Multiphase condensation may not always happen in the ICM and CGM when
the background min$(t_{\rm TI}/t_{\rm ff})$ is high. Recent X-ray observations like \cite{lakhchaur17} discuss the existence of red nugget galaxies, which
are the building blocks for compact, massive galaxies at the current time and which have no recent star formation. One such system, Mrk 1216, has very short cooling times and small core entropy, which suggests that gas in the galactic core should condense out of the medium. But, with deeper gravitational potential, the $t_{\rm cool}/t_{\rm ff} \approx 20$ is somewhat high and it doesn't show any signatures of recent condensation. Such systems, unlike the standard cool-core clusters (\citealt{pulido17}), can distinguish $t_{\rm cool}/t_{\rm ff}$ models from those based on just $t_{\rm cool}$ or the core entropy. 
Since halos grow hierarchically due to gravity, the $t_{\rm ff}$ profiles are rather similar for typical halos. But these galaxies with unusual growth histories can be good testbeds to understand multiphase condensation in halos.

\subsubsection{The condensation curve with $t_{\rm TI}/t_{\rm BV}$}
\label{sec:tBV}
\begin{figure*}
\includegraphics[width=0.9\textwidth]{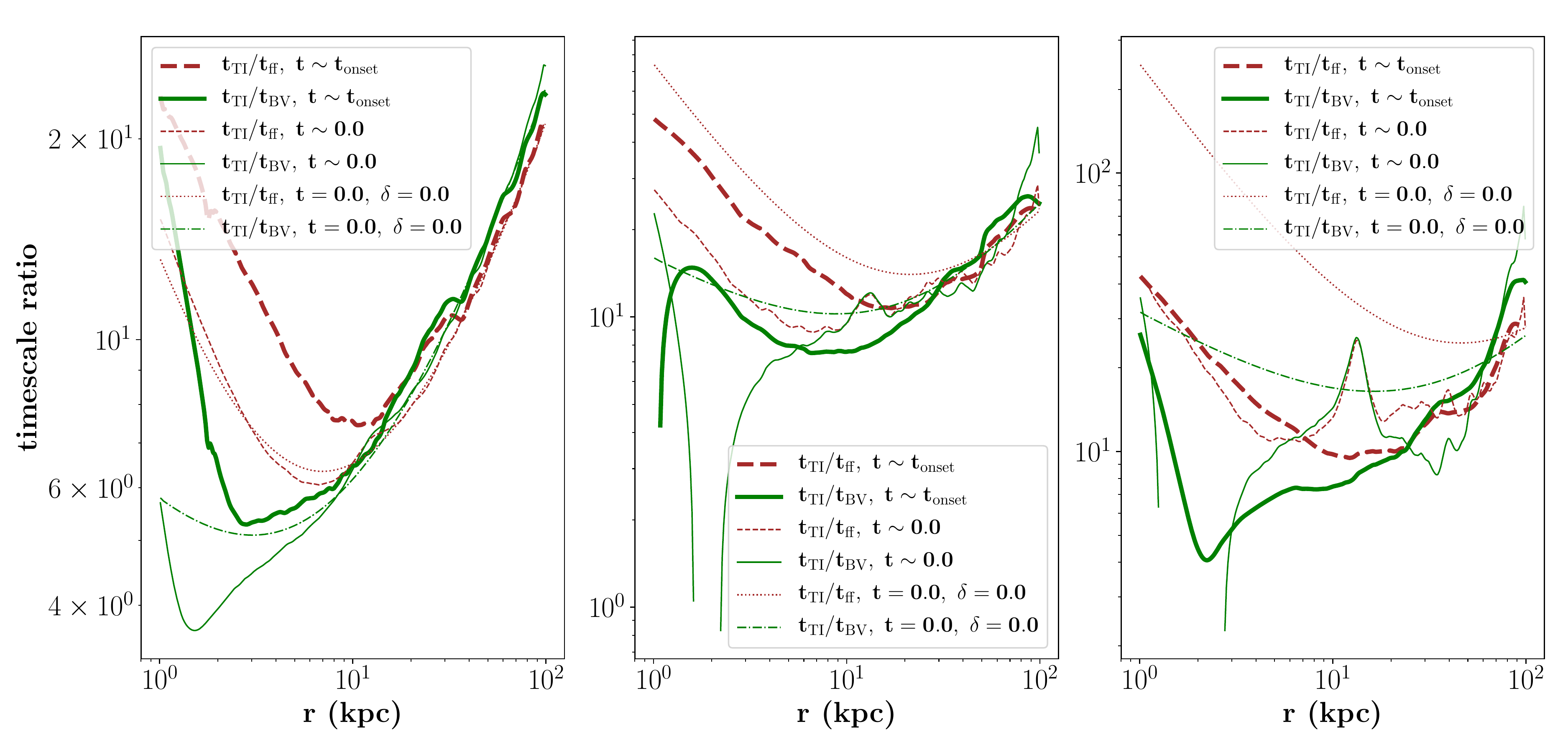}
\caption{The timescale ratios, $t_{\rm TI}/t_{\rm ff}$ and $t_{\rm TI}/t_{\rm BV}$, with radius for three different initial min($t_{\rm TI}/t_{\rm ff}$) ($6,14,25$, without perturbations, from left to right; the corresponding min[$t_{\rm TI}/t_{\rm BV}$] are $5, 10, 16$) with NFW+BCG runs from the condensation curve in Figure \ref{fig:fig6} ($\delta_{\rm max} \sim 0.001,1, 3$) and Figure \ref{fig:fig11} ($\delta_{\rm rms} \sim 0.001, 0.3, 0.6$). All the radial profiles (except ones with $\delta = 0.0$) are time-averaged within a 0.5 Gyr time window. We use a 5th order polynomial fit to smoothen the entropy profile and then take the derivative to obtain $t_{\rm BV}$. Notice that the profiles with perturbations have regions where $t_{\rm BV}$ is not defined as the background entropy gradient is negative. The range of min($t_{\rm TI}/t_{\rm ff}$) close to the onset is in the range 7-10 and  min($t_{\rm TI}/t_{\rm BV}$) is in the range 4-7.}\label{fig:fig13}
\end{figure*}

An important timescale, which characterizes the linear response of isobaric density/entropy perturbations in a stratified atmosphere, is the Brunt-V\"ais\"al\"a timescale defined as
\be
\label{eq:tBV}
t_{\rm BV} \equiv \left[ \frac{g}{\gamma}\frac{d}{dr} \ln \left( \frac{p}{\rho^\gamma}\right) \right]^{-1/2},
\ee
where we use the background pressure and density profiles. It is useful to see how the condensation curve looks like if we use $t_{\rm TI}/t_{\rm BV}$ instead of $t_{\rm TI}/t_{\rm ff}$ to quantify
the susceptibility to condensation. Figure \ref{fig:fig13} shows the timescale ratios, $t_{\rm TI}/t_{\rm BV}$ and $t_{\rm TI}/t_{\rm ff}$, for different cases (at $t\approx 0$ with and without perturbations, and at the onset of multiphase condensation, $t\approx t_{\rm onset}$). For one, it is straightforward to calculate $t_{\rm ff}$ (and hence $t_{\rm TI}/t_{\rm ff}$) unlike $t_{\rm BV}$ because the 
angle-averaged entropy profiles with perturbations have sharp jumps and negative radial derivatives at some points (especially where the atmosphere is close to isentropic). 
Since min($t_{\rm TI}/t_{\rm BV}$) appears to be a factor of $\sim 2$
lower than min($t _{\rm TI}/t_{\rm ff}$) in Figure \ref{fig:fig13} at the onset of multiphase condensation, we expect the condensation curve (Fig. \ref{fig:fig6}) based on min($t_{\rm TI}/t_{\rm BV}$)  to be shifted downward by a factor of $\sim 2$ compared to the curve with min($t _{\rm TI}/t_{\rm ff}$).


Figure \ref{fig:fig13} also shows that even if the initial background has a large $t_{\rm TI}/t_{\rm ff}$, the perturbation level (required at the threshold of condensation) is such that the $t_{\rm TI}/t_{\rm ff}$ remains around $\sim 10$ (see the thick dashed lines), which is consistent with our phenomenological model presented in section \ref{sec:condcrv} that argues that an overdense blob leads to
multiphase gas if its $t_{\rm TI}/t_{\rm ff} \lesssim 10$, even if the background has a higher ratio.

\subsection{Role of entropy gradient in condensation}
\cite{voitglobal17} gives significant emphasis on the role of entropy gradient in regulating feedback. According to their model, cold gas can condense out in two regions
via different mechanisms: in the central isentropic zone due to thermal instability and due to uplift in the outer zones with a steeper entropy profile.
The authors argue that the power-law entropy zone plays a crucial role in the cool core feedback cycles. We confirm that multiphase condensation is harder and less widespread with steeper entropy profile. Fast buoyancy oscillations are expected to damp the linear growth of multiphase condensation once the perturbations become larger. In the absence of buoyancy oscillations, condensation should be easy. But contrary to the expectations from linear theory, the constant entropy models require a smaller value of the threshold min$(t_{\rm TI}/t_{\rm ff})$ for condensation compared to the core+power-law profile (making the onset of condensation appear more difficult; see Fig. \ref{fig:fig6}). The reason for this is the long cooling time for this case relative to the run-time (the cooling time at the location of min$[t_{\rm TI}/t_{\rm ff}]$ is much shorter for a rising entropy profile). As the condensation sets in, it is relatively more widespread in the constant entropy model. However, buoyancy oscillations and the absence of multiphase gas are not always coincident in this case across the entire run-time. This implies that strong buoyancy oscillations are sufficient but not always necessary to prevent multiphase condensation. 

\begin{figure*}
\includegraphics[width=0.9\textwidth]{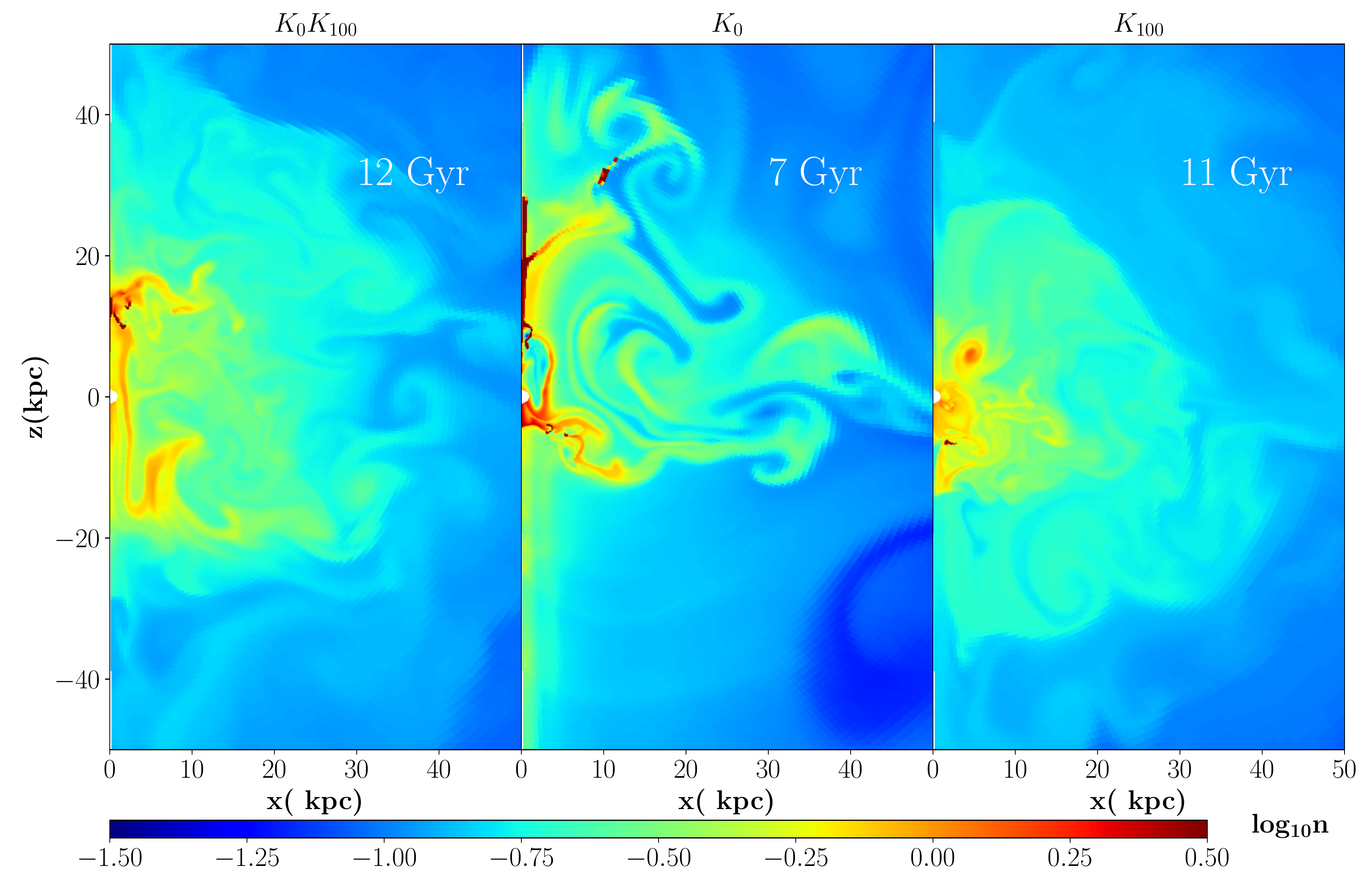}
\caption{The number density snapshots for the three different entropy profiles $K_0K_{100}$, $K_0$ and $K_{100}$, for $0.65 q_{\rm thresh}$ at one of the times when the cores have extended cold gas. For $K_0$ gas condenses out along dense filaments distributed over a much larger region compared to the other two cases. In axisymmetric simulations using spherical
coordinates there is a tendency of cold gas to accumulate close to the poles for some time, as seen here. However, multiphase gas originates away from the poles and our key results are robust.}\label{fig:fig12}
\end{figure*}

\subsubsection{Implications of shallower entropy profile in the CGM}
A large fraction of the CGM in smaller mass halos is expected to be within the location of min$(t_{\rm cool}/t_{\rm ff})$ (or in other words, expected to have large isentropic cores;
\citealt{sharma12b}, \citealt{mallerbullock04}), and hence multiphase condensation is expected to be more widespread. For our runs with constant entropy throughout the medium, the location of min$(t_{\rm cool}/t_{\rm ff})$ is pushed almost up to the outer boundary.
These runs mimic a CGM-like atmosphere with a large radial extent of the central core. The threshold min$(t_{\rm cool}/t_{\rm ff})$ required for multiphase condensation in these runs, is, in fact, less than the ones which have a power-law entropy profile in the outskirts. But, we observe that once condensation is triggered, there is a large amount of gas cooling out of the background medium and over a large radial extent (see Figure \ref{fig:fig7} and Figure \ref{fig:fig10}).
This may explain the observation  (\citealt{tumlinson17}) of cold gas along most lines of sight through the CGM. However, for a CGM, the background medium is not necessarily in global thermal balance, particularly for reasonably small halo masses. Thus our results should be taken to be qualitative but robust indicators. 

In our runs, the medium with constant entropy has relatively more condensed gas and is clearly more disrupted than the other two entropy profiles if we see the cores at one of the times of peak activity. Figure \ref{fig:fig12} shows the number density maps for all the cases at one of these times, which are not necessarily coincident. For $K_0$, the low-density hot bubbles generated in the central region (produced because of our heating prescription) can move out up to large radii freely, due to the lack of buoyant oscillations. Consequently, such bubbles grow and perturb the gas at large radii. We see distinct filaments spreading up to $\sim 40-50$ kpc and gas condenses out along these. For the core+power-law entropy profile, cooling and heating are confined within smaller radii. In the pure power-law profile, the cooling and disrupted phase of the gas remains more tightly confined than the former. 

For the CGM it has been theoretically predicted that the virial shock can be unstable for significantly lower mass halos (\citealt{birnboimdekel2003}). Dense streams can enter the halo (\citealt{2005MNRAS_keres}) providing large overdensities that may seed multiphase condensation. 
Observations of metal-rich cold gas in the outskirts by COS-Halos observations seem to imply that the cold clumps seen along all lines of sight are recycled by feedback (\citealt{2013ApJ_werk}, \citealt{2016MNRAS_ford}). However, we cannot rule out the possibility of large perturbations seeded by relatively pristine dense streams (of course the local IGM is polluted by outflows at much earlier era). Dense ram pressure stripped gas from satellite galaxies orbiting massive groups and clusters (\citealt{2019MNRAS_yun}) may seed large overdensities in the hot medium. Molecular gas has been observed to be in the stripped tail of such galaxies (\citealt{2017ApJ...839..114J}). Magnetic fields can also aid multiphase condensation by providing magnetic support against gravity and thus seeding large density perturbations (\citealt{2017arXiv171000822J}). 

\subsubsection{How do the entropy profiles evolve?}
While it is most convenient to plot the condensation curve (Fig. \ref{fig:fig6}) in terms of the background unperturbed value of min($t_{\rm TI}/t_{\rm ff}$), the background shell-averaged profiles (and 
hence $t_{\rm TI}$ and $t_{\rm BV}$) change with perturbations and in time (see Fig. \ref{fig:fig13}).
\citet{voitglobal17,2018ApJ_voit} argue that buoyancy oscillations are driven by thermal instability, and the density fluctuations saturate when the buoyancy damping rate balances the growth rate; i.e., $t_{\rm TI}\sim t_{\rm BV}$. Another possibility for saturation, especially for shallow entropy gradients, is obtained via balancing $t_{\rm TI}$ and the turbulent 
eddy turn-over time (\citealt{mccourt12,ashmeet14}). We note that our initially isentropic runs ($K_{100}$) develop a mild positive entropy gradient before the onset of condensation. It appears
unlikely that the weak entropy gradient in these cases plays a significant role. However, our results with larger entropy gradient runs strongly suggest that a steep entropy gradient suppresses condensation.

It is useful to understand the entropy evolution of different background profiles with and without heating/cooling. We perform some of the runs in Table \ref{table:listofruns2} to characterize how entropy profile is modified without heating and cooling. All these runs show a small increase in the central entropy (due to the dissipation of gravity waves) and no entropy sorting. When cooling and heating are included, core+power law and constant power-law runs ($K_0K_{100}$ and $K_0$) show the entropy becoming smaller in the central region as the lower entropy material at larger radii moves in (i.e., entropy sorting), which implies that entropy sorting is caused by cooling and heating. For a power-law profile ($K_{100}$), the central entropy increases rather than decreasing. After significant condensation, all thermal balance simulations seem to approach a similar entropy profile in the core (compare Figs. 5 \& 7 in \citealt{sharma12}), which can be understood from the tendency of such atmospheres to become marginally 
susceptible to multiphase condensation in the core.

\begin{figure*}
\includegraphics[width=0.9\textwidth]{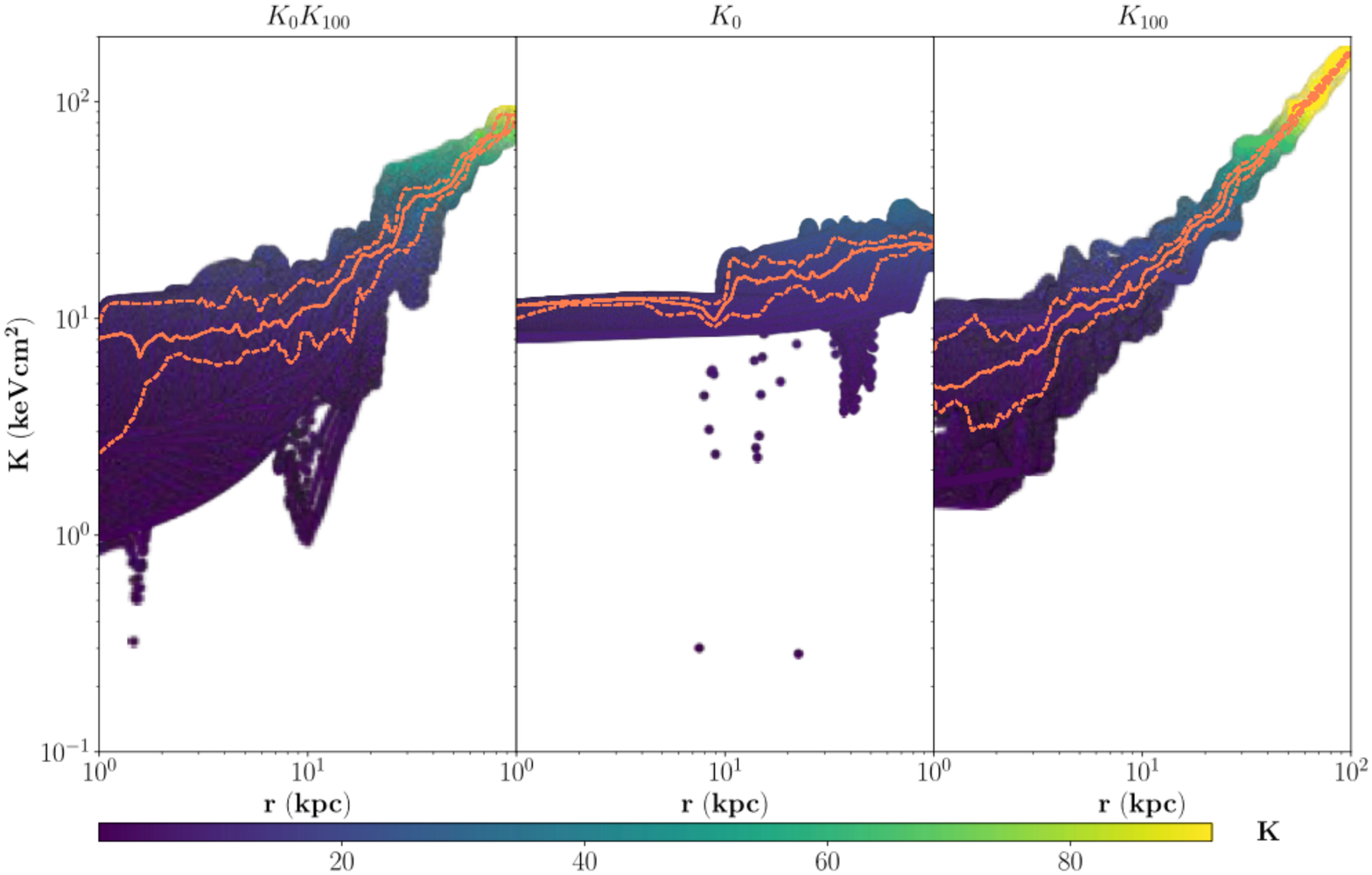}
\caption{The entropy distribution (across $\theta$-direction) of all grid points in a given radial shell for $K_0K_{100}$, $K_0$ and $K_{100}$ runs at the onset of condensation with the colorbar 
showing the entropy on a linear scale.
The coral solid lines show the median entropy profile at each radius and the coral dashed lines enclose the regions between $20-80$ percentile. }\label{fig:fig14}
\end{figure*}

Figure \ref{fig:fig14} shows the distribution of entropy in shells for the three cases ($K_0K_{100}$, $K_{100}$, $K_0$ including cooling/heating) at the onset of cold gas. The solid coral lines show the medians, and the dashed coral lines enclose the regions within $20-80$ percentile, and symbols indicate the entropy distribution of individual grid cells. This figure demonstrates several important aspects of entropy evolution. Firstly, the extreme low-entropy region is spread out till very large radii for the flat entropy run ($K_0$), somewhat less in $K_0K_{100}$ and the least in $K_{100}$. This radial extent of cold and hot gas is within the location of min$(t_{\rm TI}/t_{\rm ff})$ in all the three cases. Secondly, the outskirts in $K_0$ have a larger spread in entropy compared to the central regions, contrary to $K_0K_{100}$ and $K_{100}$. This implies that despite the sorting of entropy in the central region (enhanced entropy gradient) for $K_0$, multiphase condensation starts in the outskirts.  Thirdly, the entropy is typically higher in $K_0$ than the central regions of $K_0K_{100}$ and $K_{100}$ despite the condensation of a larger amount of cold gas in the former. This sums up how it is easier to obtain significant cold gas in a constant entropy environment similar to the CGMs in which a larger volume is susceptible to multiphase condensation.

\subsection{Role of outflows and bubbles in condensation}
AGN jets can promote multiphase condensation in two ways: their trails entraining gas from the core and
in case of supersonic jets, impinging into the medium and compressing regions in which density is enhanced. In our setup, there are overdense
regions seeded already and the gentle outflows that we have can simply entrain the gas. 
We inject low jet powers (and thermal balance in shells) so perturbations can become denser due to entrainment and there is an onset of condensation with a smaller amplitude than what is predicted by the background condensation curve. With an injection of higher jet powers, it is relatively easier to obtain cold gas, particularly for a small background $t_{\rm cool}/t_{\rm ff}$ (comparing the lime-green and golden right triangles
in Figure \ref{fig:fig6}). However, it is worth mentioning that extremely powerful jets heat up the core on an average and the core has lower density for very long times (\citealt{prasad15}). In such a case, the total amount of cold gas (in an average sense) decreases in the system. In our setup, condensation becomes difficult for supersonic jets due to overheating.

Contrary to the jets, we observe that our buoyant bubbles have a negative impact on condensation. The bubble is disrupted after moving a small
distance away from the center. But it prevents condensation particularly for small values of background min$(t_{\rm cool}/t_{\rm ff})$, reducing this threshold to lower values for same amplitudes, compared to what is expected from the condensation curve. \cite{revaz08} show condensation occurring in the wake of large scale bubbles in their ICM simulations.
However, the background ICM in these simulations, is not in thermal equilibrium, with a cooling flow occurring in a few hundred megayears. Without any heating source,
the overdensities in the wake of the bubble can condense out easily given enough time. In fact, these simulations were run for only $600$ Myrs and the minimum cooling time was around $400$ Myrs. 
The cooling catastrophe can dominate in a few hundred Myrs for this case and most of the gas will cool down to very low temperatures even without a bubble. Our setup is qualitatively different because of the presence of heating (which we implement in the form of idealized thermally balanced shells). 

Thus while jets are conducive to condensation of reasonably large density perturbations, bubbles suppress the condensation. However, the magnitude of the effect is mild and does not affect the background condensation curve to a great extent. A thorough exploration of the entire parameter space of bubbles and jets is beyond the scope of the present work.


\section{Conclusions}
\label{sec:last}
We carry out a suite of simulations to explore the factors that affect multiphase condensation in the CGM. Earlier simulations (\citealt{mccourt12}, \citealt{sharma12}) 
considered only small amplitude perturbations in the intracluster medium in hydrostatic and global thermal balance. They find that min$(t_{\rm cool}/t_{\rm ff})$ in the background medium 
plays a pivotal role
in determining whether multiphase condensation happens. While \citealt{sharma12} put a threshold ($\approx 10$) for cold gas to form, which is corroborated by subsequent observations
and simulations, \cite{choudhury16} explored the possibility of slightly higher values of threshold min$(t_{\rm cool}/t_{\rm ff})$  for somewhat different potentials and find that it can vary within a 
factor of $2$. Recent
observations hint at multiphase medium for background $t_{\rm cool}/t_{\rm ff}$ as high as $\approx 25$ (\citealt{hogan17}). This partly motivates our investigation of the possible
conditions under which a medium, with large background min$(t_{\rm cool}/t_{\rm ff})$, can be susceptible to multiphase condensation. Our numerical experiments reveal that theoretically, it is always possible to condense gas out of the hot medium if the initial density inhomogeneities are large enough (\citealt{ps05}, \citealt{ashmeet14}). Such large density perturbations can be due to dense regions formed around low-density AGN-driven bubbles, stripping of dense gas from galaxies moving through the circumgalactic medium, cosmological cold filaments breaking up as they enter the virial radius, etc. 

The main conclusions of our work are the following:
\begin{itemize}
 \item {\bf The condensation curve:} We introduce a condensation curve in the min$(t_{\rm TI}/t_{\rm ff})-\delta_{\rm max}$ plane which defines the regime in which multiphase condensation occurs in the ICM/CGM. This curve also shows that below an amplitude $\delta_{\rm max} \approx 1$ ($\delta_{{\rm min}(t_{\rm cool}/t_{\rm ff})} < \delta_{\rm max}$), it is the $t_{\rm TI}/t_{\rm ff}$ ratio of the background atmosphere which
 governs condensation. For large perturbation amplitudes, the condensation curve rises steeply, implying that the background $t_{\rm TI}/t_{\rm ff}$
 has a relatively small role to play. On the contrary, the $t_{\rm TI}/t_{\rm ff}$ {\it of the condensing blob} is what plays a crucial role in determining multiphase condensation. The condensation curve corresponds to the threshold $t_{\rm cool}/t_{\rm ff}$ {\it of the blob} $\sim 10$ irrespective of the background (see section \ref{sec:condcrv} and eq \ref{eq:blob}). We anticipate the condensation curve to shift downward by a factor of $\sim 2$, if defined in terms of min$(t_{\rm TI}/t_{\rm BV})$ (see section \ref{sec:tBV}). It will be useful to create a similar condensation curve relative to the turbulent 
 velocity (say in the min[$t_{\rm TI}/t_{\rm ff}$]-turbulent velocity plane) because turbulence can not only produce density fluctuations (\citealt{mohapatra2019,zhuravleva2014}) but can also directly affect the physics of multiphase 
 condensation (\citealt{2018ApJ_voit}).
 
 We delineate the susceptibility of the medium to produce multiphase condensation but do not quantify in detail the amount of cold gas formed. However, the zone of condensation is below this curve and we show that we get more cold gas if we move away from the condensation curve into this zone. 
 
 \item {\bf Localized perturbations slightly deviate from the curve but follow the trend:} The condensation curve is defined by labeling our runs with the background
 min$(t_{\rm TI}/t_{\rm ff})$ and the maximum value of the amplitude of density perturbations ($\delta$). However, $\delta$ and $t_{\rm cool}/t_{\rm ff}$ (or $t_{\rm TI}/t_{\rm ff}$)
 vary spatially. Local condensation depends on the local values of these parameters. In order to understand how the radial location of overdensity affects multiphase condensation, we seed perturbations only in narrow radial shells. These tests show that the amplitude of perturbations required for condensation, in case of localized perturbations, is close to what we expect from the condensation curve. 
 As one moves out from the radius of min$(t_{\rm TI}/t_{\rm ff})$, it is harder to get cold gas because the local $t_{\rm cool}/t_{\rm ff}$ is large. However, note that outside the radius of min$(t_{\rm TI}/t_{\rm ff})$, condensation in local patches (brown points in Figure \ref{fig:fig6}) follows the locus of ${\Big(\frac{t_{\rm cool}}{t_{\rm ff}}\Big)}_{\rm blob} \lesssim 10$.
 As one moves inward from the radius of min$(t_{\rm TI}/t_{\rm ff})$, condensation is relatively easier. 
 \item {\bf Effect of entropy variation on the condensation curve:} 
 We test the effect of entropy gradient by initializing some simulations with only constant entropy and some with power-law entropy.
 A constant entropy lowers the threshold min$(t_{\rm TI}/t_{\rm ff})$, implying that the onset of condensation is somewhat difficult compared to the core+power-law entropy profile. This is because the constant entropy runs have only been run for a small number of cooling times because min($t_{\rm cool}$) is the longest. However, once condensation happens in this case, it is more widespread. On the other hand, a large entropy gradient clearly inhibits condensation.
 \begin{itemize}
 \item{\bf Implications on CGM observations:}
 In a medium with constant entropy throughout, it appears mildly difficult to initiate condensation as the threshold min$(t_{\rm cool}/t_{\rm ff})$ is slightly smaller. However, for such profiles, a larger amount of gas cools out at the threshold, at much larger radii, and progressively more into the condensation zone relative to the fiducial entropy profile. The medium is disturbed up to very large radii as both hot and cold blobs move over large distances (middle panel in Figure \ref{fig:fig12}). Hence constant entropy runs have significant implications on the observations of ubiquitous multiphase gas in the CGM as smaller mass halos are expected to have shallower entropy profiles (e.g., see Figure 1 of \citealt{sharma12}). 
  \end{itemize}
 \item {\bf Effect of outflows and bubbles on the condensation curve:} We investigate how the condensation curve is modified in the presence of bubbles and outflows. In our runs, we mimic a bubble by a patch of low density compared to the background. This locally increases the $t_{\rm cool}/t_{\rm ff}$ and prevents condensation. For outflows, we inject low power jets because we balance cooling and heating in shells and do not want to overheat due to energy injection by jets. This way we can compare the runs with and without outflows under similar conditions of background thermal balance. However, due to reasonably low powers, we only have gentle subsonic winds that essentially increases the density and shortens $t_{\rm cool}$ in the core. We see in our simulations that for high jet power ($\approx 10^{42}~{\rm ergs}^{-1}$), the threshold amplitude of perturbations required is almost an order of magnitude less than what is expected from the condensation curve, with min$(t_{\rm TI}/t_{\rm ff})\approx 10$. When we decrease the jet power by a factor of $10$, the threshold amplitude for the same background is $3$ times higher. Note that this may not be a generic result for the ICM and on an average cold gas is obtained less frequently in presence of supersonic jets and AGN feedback cycles (\citealt{prasad15}) as the powerful jet events keep the core hot for a long time.  

 \end{itemize}

Lastly, we have tried to verify the condensation curve with a couple of 3D simulations. However, we could not carry out an exhaustive scan in the parameter space even for a single background profile because of the runs being immensely expensive. From our limited exploration, we see that the threshold min$(t_{\rm cool}/t_{\rm ff})$ required for condensation is close to what is obtained in 2D for one of the backgrounds. We will do detailed comparisons with 3D simulations in the future.

\section{Acknowledgement}
PPC acknowledges MPA for a long-term visiting graduate fellowship. PS acknowledges an India-Israel joint research grant (6-10/2014[IC]) and a Swarnajayanti Fellowship from the Department of Science and Technology (DST/SJF/PSA-03/2016-17). PS also thanks the Humboldt Foundation which enabled his sabbatical at MPA where this work was finished. EQ was supported in part by a Simons Investigator Award from the Simons Foundation and by NSF grant AST-1715070. We thank our referee, Mark Voit, for important suggestions to improve the draft. 
\bibliographystyle{mn2e}
\bibliography{bibtex}
\label{lastpage}
\end{document}